\documentclass[aps,prb,twocolumn,superscriptaddress,floatfix,amsmath]{revtex4}

\usepackage{graphicx}
\usepackage{dcolumn}
\usepackage{bm}
\usepackage{color}
\usepackage{amsmath}
\usepackage{amssymb}
\usepackage{ulem}

\renewcommand{\vec}[1]{\boldsymbol{#1}}

\usepackage{dcolumn}
\usepackage{subfigure}
\newcommand{\ket}[1]{\ensuremath{\left| #1 \right\rangle}}

\newcommand{\vect}[1]{\ensuremath{\overrightarrow{#1}}}

\newcommand{\moy}[1]{\ensuremath{\left\langle #1 \right\rangle}}

\newcommand{\f}[2]{{\ensuremath{\mathchoice%
        {\dfrac{#1}{#2}}
        {\dfrac{#1}{#2}}
        {\frac{#1}{#2}}
        {\frac{#1}{#2}}
        }}}

\newcommand{\eqn}[1]{\vspace{-0.cm}\begin{equation}
#1
\end{equation}}

\renewcommand{\=}{\,=\,}
\newcommand{\+}{\,+\,}
\renewcommand{\-}{\,-\,}
\newcommand{\vv}{\vspace{-0.5cm}}

\newcommand{\trip}[3]{\ensuremath{\left[T_{#1},\left[T_{#2},T_{#3}\right]\right]}}

\graphicspath{{../}}

\begin{document}


\title{\textbf{Low energy theory of the $t-t^{\prime}-t^{\prime\prime}-U$ Hubbard Model at half-filling:
interaction strengths in cuprate
superconductors and an effective spin-only description of La$_2$CuO$_4$.}}

\author{J.-Y. P. Delannoy}
\affiliation{Universit\'{e} de Lyon,  Laboratoire de Physique, CNRS, \'Ecole normale
sup\'erieure de Lyon, 46 All\'ee d'Italie, 69364 Lyon cedex 07,
France.}
\affiliation{Department of Physics and Astronomy, University of Waterloo, Ontario, N2L 3G1,
Canada}

\author{M. J. P. Gingras}
\affiliation{Department of Physics and Astronomy, University of Waterloo, Ontario, N2L 3G1,
Canada}

\affiliation{Department of Physics and Astronomy, University of Canterbury, Private Bag
4800, Christchurch, New Zealand}


\author{P. C. W. Holdsworth}
\affiliation{Universit\'{e} de Lyon,  Laboratoire de Physique, CNRS, \'Ecole normale
sup\'erieure de Lyon, 46 All\'ee d'Italie, 69364 Lyon cedex 07,
France.}

\author{A.-M. S. Tremblay}

\affiliation{D\'epartement de Physique and RQMP, Universit\'e de
Sherbrooke, Sherbrooke, Qu\'ebec, J1K 2R1, Canada.}

\date{\today}

\begin{abstract}

Spin-only descriptions of the half-filled
one-band Hubbard model are relevant for a wide range of Mott insulators.
In addition to the usual Heisenberg exchange, many new types of interactions,
including ring exchange, appear in the effective Hamiltonian in
the intermediate coupling regime.   In order
to improve on the quantitative description of magnetic excitations
in the insulating antiferromagnetic phase of copper-oxide
(cuprate) materials, and to be consistent with band structure
calculations and photoemission experiments on these systems,  we
include second and third neighbor hopping parameters, $t^{\prime}$ and
$t^{\prime\prime}$, into the Hubbard Hamiltonian. A unitary
transformation method is used to find systematically the effective
 Hamiltonian and any operator in the spin-only representation.
The results include all closed, four
hop electronic pathways in the canonical transformation. The
method generates many ring exchange
terms that play an important role in the comparison with
experiments on La$_2$CuO$_4$. Performing a spin wave analysis, we
calculate the magnon dispersion as a function of $U,t,t^{\prime}$ and
$t^{\prime\prime}$. The four parameters are estimated  by fitting the magnon
dispersion to the experimental results of Coldea {\it et al.}
[Phys. Rev. Lett. {\bf 86}, 5377, {2001}] for La$_2$CuO$_4$.  The
ring exchange terms are found essential, in particular to
determine the relative sign of $t$' and $t^{\prime\prime}$, with the values
found in
good agreement with independent theoretical and
experimental estimates for other members of the cuprate family.
The zero temperature sublattice magnetization is calculated using
these parameters and also found to be in good agreement with the
experimental value estimated by Lee {\it et al.} [Phys. Rev. B
{\bf 60}, 3643 (1999)]. We find a value of the interaction strength $U \simeq 8t$ consistent with Mott insulating behavior.

\end{abstract}

\maketitle

\section{Introduction}

High temperature superconductors have challenged almost every
traditional  concept and method of condensed matter theory. Ill
understood issues concerning Fermi liquids and quantum critical
behavior for example, may be intimately related to the problem of
superconductivity itself. From the very beginning, it  has therefore
appeared important to develop a better understanding of quantum magnetism in
general, given that the parent compounds of high-temperature
superconductors are insulating antiferromagnets. For example,
although these compounds are N\'eel ordered, it was proposed early
on that, in two dimensions, the ground state of the Heisenberg
antiferromagnet might be very close to being a resonating-valence
bond state rather than a N\'eel ordered state~\cite{Anderson3}.
This turns out not to be true, but it took some time to establish
numerically that the ground state is indeed
ordered~\cite{Reger:1988,hirsch:1989}, in agreement with
experiment~\cite{Shirane:1987,Endoh:1988}.

Issues relating to the quantum magnetism of the copper-oxide
materials (cuprates) have resurfaced in the last few years.
Detailed neutron scattering experiments on the parent
high-temperature superconductor La$_2$CuO$_4$ illustrate that
the question of which model
best describes these compounds~\cite{coldea01}
can now be addressed with precision. For example, while
the magnetic properties of Copper Deuteroformate Tetradeuterate
can be described in detail using the Heisenberg
model~\cite{Aeppli:2001}, La$_2$CuO$_4$ exhibits clear
deviations~\cite{coldea01} from it.

Specifically, the observed spin wave dispersion along the
antiferromagnetic Brillouin zone boundary is not predicted by the Heisenberg model.
One can in principle
account for these differences by introducing phenomenological
two-spin and multi-spin exchange constants into the spin
Hamiltonian. These spin Hamiltonians can be deduced, as shown by
Dirac~\cite{Dirac:livre,Dirac:1929,Roger:2005}, from general
considerations of symmetry and permutation operators and the value
of the exchange constants can be fixed by fitting to the
experimental data. However, in our opinion, a more interesting
approach
 is that taken by Coldea {\it et al.}~\cite{coldea01}.
The authors of
 Ref.~[\onlinecite{coldea01}]
 take the point of view, which we share, that the one-band
Hubbard model is a more fundamental starting point for the
description of the copper oxygen planes in these materials if one
wants to connect the magnetism at the microscopic level with
electronic correlation effects. Although in this paper we
concentrate on half filling, it  may
 also be valid on moving
away from half-filling, into the region where the fermions
eventually pair to give high-temperature superconductivity.

The one band Hubbard Hamiltonian $H_{\rm H}$ is one of the simplest
lattice models of interacting
electron that admits a ``rich'' phenomenology resembling that
of the cuprates.  A key feature of this model is that it does
describe both the insulating and metallic phases and the (Mott) transition
between them. In interaction-induced insulators, the band is half-filled
and insulating behavior occurs because of interactions, not because of
band folding induced by long-range order. However, in two-dimensions
a half-filled band can be insulating because of a pseudogap induced
by fluctuating antiferromagnetism, or because of the Mott phenomenon.
In the former case (weak coupling), the pseudogap appears when the
thermal de Broglie wavelength $\xi_{\rm th}$ is less than the antiferromagnetic
 correlation length~\cite{Vilk:1997} $\xi$, while in the latter Mott insulating case
(strong coupling), the system is insulating even when this
condition is not satisfied. In La$_2$CuO$_4$ neutron measurements have been
done only in the regime where $\xi > \xi_{\rm th}$, so 
that antiferromagnetic correlations could be the
source of the insulating behavior.
In a very recent paper~\cite{Comanac}
Comanac {\it et al.} used {\it single-site}
dynamical mean-field theory (DMFT) to calculate the optical
conductivity of La$_2$CuO$_4$.
The authors of Ref.~[\onlinecite{Comanac}] found
that La$_2$CuO$_4$, or at least its optical conductivity, is best
parametrized by a $U/t$ smaller than that necessary for the
Mott transition in single-site DMFT. Consequently, the authors of
Ref.~[\onlinecite{Comanac}] argue that correlations in the cuprates may not be
as strong as generally believed and that these materials may not be
Mott insulators but rather, that the development of antiferromagnetism
is necessary to drive the the insulating state.  On the other hand,
{\it cluster}-DMFT calculations, which are more accurate in two dimensions,
find that the inclusion of short-range
antiferromagnetic correlations, which is possible in {\it multi-site}
calculations, reduce considerably the critical $(U/t)_{\rm Mott}$
necessary for the Mott transition compared with single-site
DMFT~\cite{Imada:2007, Park-cond-matt, Gull:2008, Ferrero:2008}.

This leads to one of the key questions in this field. Do the
strange properties observed in underdoped cuprates at finite
temerature arise from interaction-induced
localisation (Mott insulator) or from competing phases?
If interactions are not strong enough to lead to a Mott insulator,
competing phases arising at weak coupling are the only remaining possibility.

To find parameters appropriate for high-temperature superconductors, Coldea {\it et al.} used earlier theoretical
results~\cite{taka,mac} relating the parameters of the Hubbard
model, hopping energy $t$ and on-site interaction $U$, to the
exchange constants in an effective spin-Hamiltonian. The
parameters in the Hubbard model were deduced by fitting the
resulting effective low-energy spin theory to the spin-wave
dispersion relation. This is quite a remarkable result. Indeed,
had the Heisenberg limit accurately described the Hubbard model,
only the value of $t^2/U$ could have been deduced from neutron
experiments. The appearance of corrections to the Heisenberg model
to higher order in powers of $t/U$ allows one to obtain $t$ and $U$
separately, the former being an effective electronic band quantity.


\begin{figure}[ht]
\includegraphics[width=6.5cm]{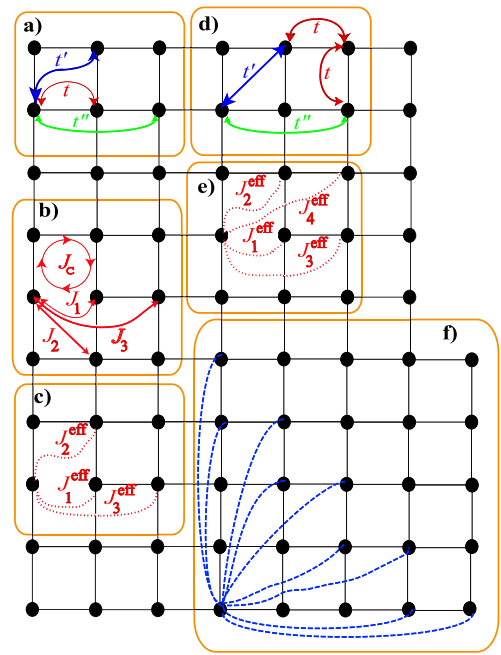}
\caption{(Color online) Hopping processes and resulting spin interactions.
{\bf a)} shows the different hopping processes characterized by parameters
$t$, $t^{\prime}$ and $t^{\prime\prime}$ in the Hubbard model considered in this paper.
$t$, $t^{\prime}$ and $t^{\prime\prime}$ are hopping parameters between
first, second and third nearest neighbors, respectively.
{\bf b)} At half-filling, and when $t^{\prime}=t^{\prime\prime}=0$, canonical perturbation theory
leads to order $t^4/U^3$ to an effective spin-1/2 Hamiltonian, $H_{\rm s}^{(4)}$,
characterized by first ($J_1$), second ($J_2$) and third ($J_3$)
nearest neighbor exchange interactions
as well as a 4-spin ring (cyclic) exchange interaction with coupling strength $J_c$.
{\bf c)} In a large $S$ expansion, $H_{\rm s}^{(4)}$ can be recast to order $1/S$
as an effective spin Hamiltonian which only involves bilinear (pairwise)
spin-spin exchange interactions between first ($J_1^{\rm eff}$), second ($J_2^{\rm eff}$)
and third ($J_3^{\rm eff}$) nearest neighbors.  To order $1/S$, the effect of the ring exchange
of strength $J_c$ merely renormalizes the first and second nearest-neighbor exchange
as discussed in Ref.~[\onlinecite{coldea01}].
{\bf d)} illustrates an example of a four hops (ring exchange)
electronic process that involves $t^{\prime}$ and $t^{\prime\prime}$ and contribute to order
$t^2t^{\prime} t^{\prime\prime}/U^3$ to the spin Hamiltonian $H_{\rm s}^{(4)}$.
{\bf e)} shows that the ring exchange term illustrated in {\bf d)} introduces in
the $1/S$ approximation of $H_{\rm s}^{(\rm eff}$ a fourth nearest neighbor exchange,
$J_4^{\rm eff}$ as well as renormalize the
$J_1^{\rm eff}$, $J_2^{\rm eff}$ and $J_3^{\rm eff}$ of {\bf c)}.
{\bf f)} illustrates all the additional $J_{n\ge 4}^{\rm eff}$ exchanges beyond
those shown in {\bf c)} and which are
generated by all the hopping processes involving an allowed combination
of $t$, $t^{\prime}$ and $t^{\prime\prime}$ to order $1/U^3$.
}
\label{hops}
\end{figure}

However, to be consistent with a general message provided by Angle
Resolved Photoemission Spectroscopy
(ARPES)~\cite{Yoshida:2005,Ino:1999,Ino:2002,Damascelli} experiments, optical
experiments~\cite{Schuttler} and with results from band structure
calculations~\cite{pickett:1989,Andersen:2001} on a wide variety
of quasi two-dimensional cuprate materials, it is necessary to
include in the Hubbard model both second ($t^{\prime}$) and third ($t^{\prime\prime}$)
nearest-neighbor hopping, as they have been found to
have sizable values in all the above studies (see
Fig.~\ref{hops}). In this paper we thus address the following
questions: What is the effect of these additional hopping
constants on the magnetic properties of the cuprates, and on
La$_2$CuO$_4$ in particular? Can their ``effective'' values and that of $U$ within
the N\'eel order phase be determined by comparing experimental
spin wave dispersion data with theoretical calculations?



As a first step to address the question of the strength of
$t$, $t^{\prime}$, $t^{\prime\prime}$ and $U$ in La$_2$CuO$_4$,
we incorporate the effects of these extra hopping
constants into a one-band Hubbard model from which we derive an
effective spin-only Hamiltonian description of La$_2$CuO$_4$. As
we shall see, the effects of $t^{\prime}$ and $t^{\prime\prime}$ are rather subtle and
can even sometimes compensate each other. It is only when these
hopping terms have been included in the derivation of the
effective spin Hamiltonian, with all the subsequent new ring
exchanges, that one can extract more meaningful values of
$t-t^{\prime}-t^{\prime\prime}-U$ from comparisons of theory with inelastic neutron
scattering experiments. In particular, the values of $t$ and $U$
obtained by Coldea {\it et al.}~\cite{coldea01} place the underlying
Hubbard model in the intermediate coupling regime where the band
width and interaction energies are comparable and hence in the
regime where one might expect an insulator to metal transition.
Our more detailed analysis leads to a larger values of $U$,
repositioning material
relevant model parameters more clearly within the Mott
insulating regime, as we will discuss more thouroughly at the end of this paper.


Although we analyze experiments on La$_2$CuO$_4$, providing a
specific materials context to our study, we believe that the
parameters of the one-band Hubbard model that we extract below
should be characteristic of the two-dimensional copper-oxygen
planes of cuprate superconductors in general: specifically,
band structure
calculations do reveal variations of band parameters from one
compound to the next~\cite{Andersen:2001,Markiewicz}, but overall,
the band structure of CuO$_2$ planes is quite similar from one
compound to the next~\cite{Kim:1998}. Even away from half filling,
whether for hole or electron doped materials, the one-band Hubbard
model for CuO$_2$ planes seems to contain much of the Physics of
high-temperature
superconductors~\cite{Anderson:2006,Tremblay:2006}. Our
calculations should therefore have wider ranging applications than
the detailed comparison with experiments on pure La$_2$CuO$_4$
presented below.

In the next two subsections of this Introduction, we describe in
more detail the central issues and key results of this paper.
There are two main parts to our work. In the first part, Section
\ref{Effective-H}, we derive the effective Hamiltonian to order
$1/U^3$, including $t-t^{\prime}-t^{\prime\prime}$. This derivation is exact and
completely general. In the second part, Section III, we determine
the spin wave energies, taking into account the  quantum
correction to the classical frequencies through the spin wave
velocity
renormalization factor, $Z_c(\vec k)$, to lowest order in $1/S$. We
then determine a parameter set $\{t,t^{\prime},t^{\prime\prime},U\}$ by fitting the
experimental magnon excitation spectrum of
Ref.~[\onlinecite{coldea01}]. The resulting set is in good
agreement with values from ARPES experiments and band structure
calculations on various cuprates. Further, we calculate the zero
temperature staggered magnetization order parameter using the set
$\{t,t^{\prime},t^{\prime\prime},U\}$ and compare it with experimental values. A more
complete solution of the spin dynamics for the effective spin
Hamiltonian derived herein, including magnon-magnon interactions,
would most likely require the use of numerical
techniques~\cite{Melko:2005}, since it would be an arduous task to
include in the self-energy corrections the multitude of magnon
creation and annihilation terms present beyond the non-interacting
approximation.

\subsection{Spin-only description of Hubbard model and La$_2$CuO$_4$}

In many strongly correlated quantum mechanical systems, the
separation of energy scales allows the development of a low energy
effective theory, through the integration over the high energy
degrees of freedom. In this context, spin-only descriptions of
strongly-correlated electron systems are an excellent example.
Here, in the limit of strong electron-electron interaction,
elimination of the states with one or more doubly occupied sites
reduces exponentially the dimension of the relevant low energy
Hilbert space, facilitating the theoretical and numerical
description of the problem. Through this procedure, the regime of
particular interest, where energy scales actually compete, can be
approached perturbatively, via the introduction of a small
parameter, the ratio of the kinetic to potential energy scales.
However, the price one pays for this dimensional reduction is the
generation of longer range and many particle interactions, as one
moves into this intermediate regime. The derivation of such a
theory can be achieved using different methods, leading to
distinct expressions for the effective Hamiltonian, which can be
shown to be equivalent through the application of a unitary
transformation~\cite{trem}. One of these methods, the canonical
transformation (CT)~\cite{Harris,mac,Delannoy},
which we adopt below, is a
convenient and systematic way of expanding a model like the
Hubbard model to any order in the perturbation parameter.

In the regime where $t/U$ remains a small parameter suitable for a
perturbation theory, but where one moves away from the strictly
Heisenberg limit ($t/U \rightarrow  0$), the pairwise exchange
interactions are joined by ring, or  cyclic  exchange terms that
couple more than 2 spins~\cite{mac}. Large enough ring exchange
terms have been found to drive some model systems into exotic,
intrinsically quantum mechanical ground
states~\cite{Singh-ring,Lauchli,Motrunich}. Indeed, there is
currently a rapidly growing interest in the study of effective
many-body non pairwise spin$-$spin interactions which lead to
non-trivially correlated states in quantum spin systems
~\cite{Hermele,Pujol-ice}.

In most theoretical work, the effort has so far focused on models
with only the nearest-neighbor Heisenberg exchange and ring
exchange. However, when starting from a microscopic fermionic
Hamiltonian the strength of the ring exchange coupling is
determined from the small parameter $t/U$ and is not a free
parameter, nor is it the sole higher order term arising in the
resulting spin only theory. To the same order in perturbation,
$t(t/U)^3$, the canonical transformation also generates second,
$J_2$, and third, $J_3$,  neighbor  spin-spin interactions through
processes involving four electronic hops. The ring exchange,
$J_c$, is thus only one of a set of spin interactions generated to
this order~\cite{trem,mac,Delannoy} and all should be taken into
account. We remark that in the general approach of
Dirac~\cite{Dirac:1929,Dirac:livre}, the coupling constants
multiply spin permutation operators, rather than spin operators
themselves and thus have a different definition from those here.
One finds that permutation of four spins lead to both four spin
ring exchange terms and pure two-spin exchange when written in
terms of spin operators~\cite{Roger:2005}.

Within the $t-U$ Hubbard model, the inclusion of virtual hopping
pathways of $n$ hops thus introduces further neighbor spin
couplings of order $t(t/U)^{n-1}$. For a compound with a band
structure characterized by several tight-binding parameters of
comparable value, one should consider the role of direct further
neighbor hops over length scales comparable with the nearest
neighbor pathways of length $n$, as well as the multiple hop
terms. Hence, to develop a theory accurate to order $(1/U)^3$ one
should include direct hopping to first, second and third nearest
neighbors, characterized by energy scales $t$, $t^{\prime}$ and $t^{\prime\prime}$,
respectively. These hopping processes generate second and third
neighbor Heisenberg exchange terms that compete with those
generated by the four hop processes discussed above. There is
evidence that this is the situation for the CuO$_2$ planes of
parent superconducting materials: for example, in
Sr$_2$CuO$_2$Cl$_2$, which has a half filled band, the hopping
constants $t^{\prime}$ and $t^{\prime\prime}$ are estimated to be $t^{\prime}/t\sim -0.3$,
$t^{\prime\prime}/t\sim0.2$, from comparisons of exact diagonalization with
photoemission experiments~\cite{Leung:1997}. Very similar values
are found from comparisons with other types of
calculations~\cite{Tohyama:2000} and from band structure
calculations for YBa$_2$CuO$_3$O$_7$~\cite{Andersen}. In fact
these values are typical for insulating materials with
copper-oxygen planes, including both hole and electron doped
cuprates~\cite{Kim:1998,Senechal:2004}. For La$_2$CuO$_4$, band
structure calculations~\cite{pickett:1989,Andersen:2001} suggest
smaller values of $t^{\prime}/t$ and $t^{\prime\prime}/t$. ARPES experiments are not
available for this compound but they are for the doped system
La$_{2-x}$Sr$_x$CuO$_4$ that becomes a high temperature
superconductor. According to these
experiments~\cite{Yoshida:2005,Ino:1999,Ino:2002,Damascelli},
$t^{\prime}/t$ and $t^{\prime\prime}/t$ deduced from the shape of the Fermi surface are
close to band structure values, although they are larger by about
30\% for very lightly doped compound~\cite{Yoshida:2005}. We will
return to this issue later.

Given this range of values for $t^{\prime}$ and $t^{\prime\prime}$ one could expect the
resulting exchange terms of order $t^{\prime}(t^{\prime}/U)$ or $t^{\prime\prime}(t^{\prime\prime}/U)$ to be
of similar magnitude to those of order $t(t/U)^3$. This is one of
the motivations for the work presented in this paper where we
introduce all terms in a spin-only Hamiltonian that depend on
$t,t^{\prime}$ and $t^{\prime\prime}$ and that are generated up to order $1/U^3$. We
assess their importance, making particular reference to the
magnetic properties of La$_2$CuO$_4$. We show that the canonical
transformation method \cite{trem,mac,Delannoy} can be easily
generalized for the $t-t^{\prime}-t^{\prime\prime}-U$ Hubbard model, giving a complete
spin-only description for arbitrary first, second and third
neighbor hopping, correct to order $1/U^3$.

\subsection{Spin-wave analysis and comparison with experiment.}\label{second-n}


It has been known for some time that a spin wave analysis of the
spin $1/2$ Heisenberg antiferromagnet on a square lattice, taken
to leading order in $1/S$, reproduces the zero point quantum
fluctuations of the staggered moment to a good approximation
\cite{Igarashi}. This is because the second order terms
renormalize the classical magnon frequencies but do not induce
magnon-magnon interactions and make zero contribution to the
magnetization.
In this paper, we therefore take it as a reasonable first
approximation
to calculate quantum spin fluctuations to lowest order in $1/S$ and
include the first order corrections to the classical frequencies. We have
not addressed the question as to whether other physics, in
particular magnon-magnon interaction is introduced to second order
in $1/S$.

We compare our results with the inelastic neutron scattering data
of Coldea {\it et al.}~\cite{coldea01} and with the spin wave
calculation therein where only nearest neighbor hopping $t$ was
considered. We find that including further neighbor hops allows
for a more convincing description of the neutron data, with the
set of parameters ($t,t^{\prime},t^{\prime\prime},U$) that best describe the data being
comparable to other estimates for cuprate materials. Our analysis
also gives a new estimate for the total staggered magnetic moment
which is in good quantitative agreement with the experimental
estimate~\cite{Lee}.

We now describe in more detail the main features of the data that
must be explained and how going beyond the nearest-neighbor
hopping provides a
better  model. The inelastic neutron scattering
data of Coldea {\it et al.}~\cite{coldea01} provide an extensive
description  of the magnon dispersion in La$_2$CuO$_4$ over the
whole of the $1^{\rm st}$ Brillouin zone. The analysis of
Ref.~[\onlinecite{coldea01}] clearly shows that the nearest
neighbor spin $S=1/2$ Heisenberg Hamiltonian does not reproduce
the magnon dispersion over the whole zone. Most notably, while a
spin wave analysis of the Heisenberg model to order $1/S$ gives a
flat magnon dispersion over the interval ${\bm q}_{\rm BZ} \in
[(\pi,0),(\pi/2,\pi/2)]$ (along the antiferromagnetic zone
boundary), the experimental results show
{\it negative} (downward)
dispersion over this region (see Fig.~\ref{fig_coldea}). Going
beyond $1/S$ for the Heisenberg model does not explain this
negative dispersion
feature~\cite{Sandvik-dispersion,Zheng-magnon}. Rather,
introducing interactions between magnons gives rise to a small
{\it positive} (downward) dispersion over this ${\bm q}_{\rm BZ}$
interval~\cite{Zheng-magnon,Sandvik-dispersion}. Including terms
up to $t(t/U)^3$, which generates second and third neighbor and
ring exchange interactions, does however provide the
characteristic downward dispersion, and this is what allowed
Coldea {\it et al.} to fit the experimental
data within a $1/S$ spin wave
analysis~\cite{Toader,AMT_comment_Toader}. Recent analysis of
triplon excitations in the copper oxide based ladder material
La$_4$Sr$_{10}$Cu$_{24}$O$_{41}$ also suggest that ring exchange
interactions are present and are of a similar
amplitude~\cite{Notbohm07} to those found in
Ref.~[\onlinecite{coldea01}] for
La$_2$CuO$_4$.

It is important for the rest of the paper to expand briefly on
this negative dispersion and to give an interpretation for it: the
spin-only Hamiltonian, $H_{\rm s}^{(4)}$, obtained from the
Hubbard Hamiltonian, $H_{\rm H}$, to order $t^4/U^3$, contains
first, second and third nearest-neighbor exchanges, $J_1$, $J_2$
and $J_3$, respectively, as well as a four-spin ring exchange
$J_c$ (see Eq.(\ref{SS})). Within the Holstein-Primakoff spin-wave
approximation, all terms in $H_{\rm s}^{(4)}$, including the ring
exchange, contribute
to order $1/S$ quadratic magnon creation/annihilation terms
to the resulting quadratic magnon Hamiltonian $H_{{\rm s},{\rm
quad}}^{(4)}$.
To this order, the spin wave approximation thus eliminates
 the  original  multi-spin (ring exchange)
nature of the spin Hamiltonian, $H_{\rm s}^{(4)}$.
As a result, $H_{{\rm s},{\rm
quad}}^{(4)}$ could equally well have been derived from a
different spin-only Hamiltonian, $H_{\rm s}^{\rm eff}$, containing
solely
{\it bilinear}, or pairwise, spin exchange terms only.
 $H_{\rm
s}^{\rm eff}$ would have exchange terms of the form $J_{n}^{\rm
eff} {\bm S}(0)\cdot {\bm S}({\bm r}_n)$, which couple a reference
spin ${\bm S}(0)$ to an $n-$nearest neighbor spin at ${\bm r}_n$
~\cite{Toader,AMT_comment_Toader,Toader-reply}. The relationship
between the $J_{n}^{\rm eff}$ exchange couplings of $H_{\rm
s}^{\rm eff}$ and the set $\{J_1, J_2, J_3, J_c\}$ of $H_{\rm
s}^{(4)}$ is~\cite{coldea01}: \eqn{ \left\{\begin{array}{ccc}
J_1^{\rm eff} & = & J_1 - 2J_cS^2 \\
J_2^{\rm eff} & = & J_2 - J_cS^2 \\
J_3^{\rm eff} & = & J_3
\end{array}\right\}.
\label{reno}
}

\noindent where $S=1/2$, $J_1=(4t^2/U)(1-6t^2/U^2)$,
$J_2=J_3=4t^4/U^3$ and $J_c=80t^4/U^3$, and where the convention
$J_n^{\rm eff}>0$ means antiferromagnetic exchange
~\cite{Note:Ferromagnetic}. We note from Eq.~\ref{reno} that $J_c$
leads to a ``renormalization'' of $J_2$. In particular, for the
$t-U$ Hubbard model, $J_2^{\rm eff}=-16t^4/U^3$ ($S=1/2$) is
negative for all $t/U$. The negative value of $J_2^{\rm eff}$
favors ferromagnetic correlations across the diagonal of a
plaquette and gives a downward dispersion along ${\bm q}_{\rm BZ}
\in [(\pi,0),(\pi/2,\pi/2)]$ ~\cite{coldea01}, as required. In the
limit $t/U\rightarrow 0$, one recovers the nearest-neighbor
Heisenberg $S=1/2$ Hamiltonian and the dispersion becomes flat
along the zone boundary~\cite{note:RingContribution}.

Having provided an explanation for the origin of the negative
(downward)
dispersion along ${\bm q}_{\rm BZ}$, we can now discuss the effect
on the spin wave dispersion of including second nearest-neighbor
hopping $t^{\prime}$. When passing from the Hubbard model, including $t^{\prime}$,
to a spin only model, the leading order effect of $t^{\prime}$ is to
generate an  antiferromagnetic  exchange between
second-nearest neighbors
 that modify $J_2$ in $H_{\rm s}^{(4)}$:
\eqn{\begin{array}{ccccc} J_2 & \rightarrow & J_2+4(t^{\prime})^2/U & = &
4t^4/U^3+4(t^{\prime})^2/U
\end{array}}
which in turn  modifies $J_2^{\rm eff}$ in the $1/S$
spin-wave approximation in Eq.~(\ref{reno}):
 \eqn{\begin{array}{ccccc} J_2^{\rm eff} &
\rightarrow & J_2^{\rm eff}+4(t^{\prime})^2/U & = & 4(t^{\prime})^2/U-16t^4/U^3 .
\end{array} \label{J2eff}}
\noindent So, while for $t^{\prime}=0$ the dispersion is downward for all
$t/U$, a nonzero $t^{\prime}$ competes with this trend since the second
nearest-neighbor exchange generated to order $4(t^{\prime})^2/U$, is
antiferromagnetic and frustrating. This means that a sizable
increase in $t/U$ is required to achieve a good fit to the data of
Ref.~[\onlinecite{coldea01}] when $t^{\prime}$ is included in the
spin-only Hamiltonian to this order.  The same observation has
been made from a random phase approximation
calculations~\cite{singh02:_spin_la_cu0,Peres:2002} and also from
quantum Monte Carlo calculations~\cite{Gagne}, both on the half-filled
one band Hubbard model. This evolution puts
the best fit value for $t/U$ into the intermediate coupling
regime and perhaps very close, or maybe even beyond the critical
value for a metal insulator transition. If we estimate this
critical value from the value of $U$ at which a finite gap in the
density of states persists at finite temperature, then Fig. 5 of
Ref.~[\onlinecite{Vekic:1993}] suggests
 that for $t^{\prime}=t^{\prime\prime}=0$ on the square
lattice, the critical value is $t/U \sim 1/6 = 0.166$. In the
variational cluster approximation, which overestimates the effect
of interactions, the critical value of $t/U$ is larger but studies
as a function of $|t^{\prime}|$ suggest that frustration leads to a
decrease of this value \cite{Rousseau:2006}. The same trend as a
function of frustration ($t^{\prime}$) occurs on the anisotropic
triangular lattice \cite{Kyung:2006}.
Hence, with nonzero $t^{\prime}$ included in the model and presumably at play
in the real material, it is ultimately important that successful fits
to experiments on parent insulating compounds lead to {\it smaller}
values of $t/U$ than those where $t^{\prime}=0$, contrary to what is
found in the references cited above where the third nearest neighbor
hopping $t^{\prime\prime}$ is equal to zero.

As discussed above,  the third neighbor hopping term $t^{\prime\prime}$ is
estimated to be of the same order as $t^{\prime}$ in a number of cuprate
materials and should therefore also be taken into account \cite{Andersen:2001}.
However, the above discussion on the leading effect of $t^{\prime}$ might
leave one wondering whether a perturbative spin-only Hamiltonian,
starting from a half-filled one-band Hubbard model can provide  a
quantitative microscopic description of La$_2$CuO$_4$ at all. It
is this very question that has motivated us in pursuing
the work reported here, and it is why we have derived the
spin-only Hamiltonian $H_{\rm s}$ from a $t-t^{\prime}-t^{\prime\prime}-U$ Hubbard
model, including all ring exchanges and hopping processes to order
$(1/U)^3$. This procedure generates a large number of further
neighbor and ring exchange paths, a small number of which make
significant contributions. The main corrections come from terms of
order $(t^{\prime})^2/U$ and $(t^{\prime\prime})^2/U$ but we also find that ring
exchanges of order $t^{\prime} t^{\prime\prime} t^2/U^3$ are also significant.
The latter, in particular,
 determine the sign of $t^{\prime}/t^{\prime\prime}$.  Of course, the
resulting theory has now two more free parameters
(i.e. $t^{\prime}$ and $t^{\prime\prime}$)
than the original model~\cite{coldea01}
to fit the
inelastic neutron scattering data and the reader may therefore not
be surprised that we achieve a better fit than in
Ref.~[\onlinecite{coldea01}]. However, as we show in Section III,
the ensemble of parameters giving the best unbiased fit to the
data is in good agreement with the values suggested from other
sources for various cuprates~\cite{Yoshida:2005,Ino:1999,
Ino:2002,Damascelli,Markiewicz,Tohyama:2000}.

The rest of the paper is organized as follows.  In Section II we
present the method to obtain a consistent spin representation of
the original Hubbard model up to third neighbor hops.  As an
application, in Section III, we  investigate  the consequences of
applying the method to a general $t-t^{\prime}-t^{\prime\prime}-U$ model by fitting the
magnon dispersion data of Ref.~[\onlinecite{coldea01}] for
La$_2$CuO$_4$.
 Section III  also discusses the
procedure to calculate the staggered magnetization operator, and
shows that our expansion of the $t-t^{\prime}-t^{\prime\prime}-U$ model gives a value
of the sublattice magnetization in good agreement with
experiment~\cite{Lee}. We conclude the paper in Section IV. We
have included a number of appendices to assist the reader with a
few technical issues. Appendix A gives the various terms that
contribute to the effective spin Hamiltonian with arbitrary $t$,
$t^{\prime}$ and $t^{\prime\prime}$ up to order $1/U^3$. In Appendix B we show that the
renormalization factor for the magnetization, coming from charge
fluctuations~\cite{Delannoy} is the same in our approach as that
found in an extension of the mean-field Hartree-Fock method of
Ref.~[\onlinecite{schrieffer}]. Appendix C gives the ${\bf k}$
dependence of the various terms coming in the $1/S$ spin-wave
calculation. Appendix D discusses the spin-wave velocity
renormalization factor, $Z_c({\bf k})$. Appendix E comments on the
results of various constrained fits to the spin wave energies.

\section{Effective spin Hamiltonian for the $t-t^{\prime}-t^{\prime\prime}-U$ Hubbard Model}

\label{Effective-H}

A derivation of the spin-only effective theory starting from the
one band $t-U$ Hubbard model has been given by several
authors~\cite{Harris,taka,mac,Delannoy}. In this section we
investigate the effects of including new parameters $t^{\prime}$ and
$t^{\prime\prime}$, for direct hops between second and third nearest neighbors, respectively,
on the spin-only effective theory. We first recall the key steps
in the unitary transformation method for the $t-U$ Hubbard model.
We then apply the method to the $t-t^{\prime}-t^{\prime\prime}-U$ model and obtain the
modified spin-only Hamiltonian generated by all possible virtual
electronic paths up to order $1/U^3$, giving terms of order
$t^4/U^3$, $(t^{\prime})^4/U^3$, $(t^{\prime\prime})^4/U^3$, $t^2(t^{\prime})^2/U^3$,
$t^2(t^{\prime\prime})^2/U^3$, $t^2t^{\prime}t^{\prime\prime}/U^3$, and $(t^{\prime}t^{\prime\prime})^2/U^3$.

\subsection{ Derivation of the spin-only effective Hamiltonian }

We begin with a brief review of the derivation of the spin-only
Hamiltonian of the one band nearest-neighbor Hubbard Hamiltonian,
$H_{\rm H}$:
\begin{eqnarray}
H_{\rm H} &= &T + V \\ & = &  -t \sum_{i,j;\sigma}
c^\dagger_{i,\sigma}c_{j,\sigma} \+ U\sum_i
n_{i,\uparrow}n_{i,\downarrow} . \label{HH}
\end{eqnarray}
The first term is the kinetic energy term that destroys an
electron of spin $\sigma$ at nearest neighbor site $j$ and creates
one on the nearest-neighbor site $i$. The second term is the
on-site Coulomb energy $U$ for two electrons with opposite spin to
be on the same site $i$ and where
$n_{i,\sigma}=c^\dagger_{i,\sigma}c_{i,\sigma}$ is the occupation
number operator at site $i$.

As introduced by Harris {\it et al.} \cite{Harris} and developed
further by MacDonald {\it et al.} \cite{mac}, the transformation
relies on the separation of the kinetic part $T$ into three terms
that respectively increase by 1 ($T_1$), keep constant($T_0$) or
decrease by one ($T_{-1}$) the number of doubly occupied sites.

We write :
\eqn{T\,=\,-t\sum_{i,j;\sigma}c^\dagger_{i,\sigma}c_{j,\sigma}\,=\,T_1\,+\,T_0\,+\,T_{-1}}
\begin{eqnarray}
T_1 &= &-t \sum_{i,j;\sigma}n_{i,\bar{\sigma}}  c^\dagger_{i,\sigma}c_{j,\sigma} h_{j,\bar{\sigma}}\\
T_{0}&=&-t \sum_{i,j;\sigma}
h_{i,\bar{\sigma}}  c^\dagger_{i,\sigma}c_{j,\sigma} h_{j,\bar{\sigma}}  \nonumber \\
      & &+\, n_{i,\bar{\sigma}}  c^\dagger_{i,\sigma}c_{j,\sigma} n_{j,\bar{\sigma}} \\
T_{-1}&=&-t \sum_{i,j;\sigma}h_{i,\bar{\sigma}}
c^\dagger_{i,\sigma}c_{j,\sigma} n_{j,\bar{\sigma}}
\end{eqnarray}
where $\bar{\sigma}$ stands for up if ${\sigma}$ is down and for
down if $\sigma$ is up. This separation comes from multiplying the
kinetic term  on the right by $n_{i,\bar{\sigma}} +
h_{i,\bar{\sigma}} \,=\,1$ and multiplying on the left by
$n_{j,\bar{\sigma}} + h_{j,\bar{\sigma}} \,=\,1$.

Applying the unitary transformation ${\rm e}^{i{\cal S}}$ to
$H_{\rm H}$ leads to an effective Hamiltonian, $\tilde{H_{\rm s}}$,
where double occupancy is eliminated perturbatively, order by
order in ${\cal S}$. Using the relation:
\eqn{ \tilde{H}_{\rm s}^{(k)}  \,=\,{\rm e}^{i{\cal S}} H_{\rm H}
{\rm e}^{-i{\cal S}}\,=\, H_{\rm H} + \f{[i{\cal S},H_{\rm
H}]}{1!} + \f{[i{\cal S},[i{\cal S},H_{\rm H}]]}{2!} + \cdots  .
\label{commut} }
and,  defining as in Ref.~[\onlinecite{mac}]
\eqn{T^{(k)}(m_1, m_2,\cdots, m_k) \= T^{k}[m] \=
T_{m_1}T_{m_2}\cdots T_{m_k},}
${\cal S}$ is solved for, order by order. Hence, starting from
a low energy vacuum of singly occupied sites, ${\tilde{H_{\rm
s}}}$ contains no terms that create or annihilate doubly occupied
sites up to the order for which  ${\cal S}$ has been determined.
This leads to an expression for ${\tilde{H_{\rm s}}}$ to order
$t(t/U)^{3}$ (see Refs.~[\onlinecite{mac,Delannoy}]):
\begin{eqnarray}\label{H4}
\tilde {H}_{\rm s}^{(4)}  \,= & - &\f{1}{U} \, T^{(2)}(-1,1) \nonumber\\
& + &\f{1}{U^2} T^{(3)}(-1,0,1) \\
& + &\f{1}{U^3} \left(T^{(4)}(-1,1,-1,1) \- T^{(4)}(-1,0,0,1)\right. \nonumber\\\
& - &\left.\f{1}{2} T^{(4)}(-1,-1,1,1) \right),\nonumber
\end{eqnarray}
where the associated generator for the unitary transformation is:
\begin{eqnarray}
i{\cal S}^{(3)} & = & \f{1}{U}(T_1 - T_{-1}) \+ \f{1}{U^2}([T_1,T_0] - [T_0,T_{-1}])\nonumber\\
& + &  \f{1}{U^3} \left( -\trip{0}{0}{1} - \trip{0}{1}{0} - \trip{1}{1}{0}\right.\nonumber\\
& - & \left. \f{1}{4}\trip{-1}{0}{-1} \+\f{2}{3} \trip{1}{1}{-1}\right.\nonumber\\
& + & \left. \f{2}{3} \trip{-1}{1}{-1} \right)  \label{unitary}
\end{eqnarray}

\subsection{Derivation of the spin Hamiltonian.}

The effective Hamiltonian
$\tilde {H}_{\rm s}^{(4)}$
in Eq.~(\ref{H4}) is still defined in terms of fermion operators
entering the $T^{(k)}$ operators.
When focusing on the magnetic properties of the half-filled
Hubbard model, it
is convenient to recast the effective Hamiltonian
${\tilde{H_{\rm s}}}$ in Eq.~(\ref{H4}) in a spin-only notation.
For this, we use a mapping\cite{mac} between the subspace of the
Hubbard model  with singly occupied sites and the Hilbert space of
a spin $S=1/2$ system. The mapping is:
\eqn{
\begin{array}{ccc}
\mbox{Hubbard}\,\mbox{Space} & & \mbox{Spin}\,\f{1}{2}\,\mbox{Space}\\
\\
n_{i,\uparrow} = 1 & \longrightarrow & \ket{\cdots\underbrace{\uparrow}_{\mbox{site i}}\cdots}  \\
\\
n_{i,\downarrow} = 1  & \longrightarrow &
\ket{\cdots\underbrace{\downarrow}_{\mbox{site i}}\cdots}
\end{array}
\label{corres}}
The spin Hamiltonian $H_{\rm s}^{(k)}$ acting on this space is
derived from the Hamiltonian acting on the occupation number
subspace (see Ref.~\onlinecite{mac} for more details). Once it is
written in the spin $1/2$ basis it can be transformed into an
explicitly SU(2) invariant form using:
\eqn{H_{\rm s}^{(k)} \=\f{1}{2^N}
\sum_{\alpha_1,\alpha_2,\cdots,\alpha_N = 0}^3 \left(\prod_{l=1}^N
\sigma_{\alpha_l}^{(l)}\right) Tr(\sigma_{\alpha_1}^{(1)} \cdots
\sigma_{\alpha_N}^{(N)} \tilde{H^{(k)}}   ),\label{Tr}}
where $\sigma_{\alpha_p}^{(p)}$ is the  Pauli matrix for site $p$,
with electron in spin state $\alpha_p$. From this one finds the
spin-only Hamiltonian evaluated to third order in $t/U$ for the
$t-U$ Hubbard model\cite{mac,Delannoy}:

\begin{eqnarray}\label{spinHS4}
H_{\rm s}^{(4)} & = & \left(\f{4t^2}{U} - \f{24t^4}{U^3}\right) \sum_{<i,j>}\left(\vec{S_i}  \cdot \vec{S_j} \right) \nonumber\\
& + & \f{4t^4}{U^3} \sum_{<<i,j>>}\left( \vec{S_i}  \cdot \vec{S_j} \right) \nonumber\\
& + & \f{4t^4}{U^3} \sum_{<<<i,j>>>}\left( \vec{S_i}  \cdot \vec{S_j} \right)\label{SS} \\
& + & \f{80t^4}{U^3} \sum_{<i,j,k,l>}\left\{(\vect{S_i}\,\cdot\,\vect{S_j})\,(\vect{S_k}\,\cdot\,\vect{S_l})\right.\nonumber\\
&&\left.
+(\vect{S_i}\,\cdot\,\vect{S_l})\,(\vect{S_k}\,\cdot\,\vect{S_j})
\,
-\,(\vect{S_i}\,\cdot\,\vect{S_k})\,(\vect{S_j}\,\cdot\,\vect{S_l})\right\}
\nonumber
\end{eqnarray}
where $<i,j>$, $<<i,j>>$, and $<<<i,j>>>$ denote sums over first,
second and third nearest neighbors, respectively and where
$<i,j,k,l>$ identifies a square plaquette with sites $i,j,k,l$
defining
the four corners of an elementary square plaquette.

\subsection{Inclusion of $t^{\prime}$ and $t^{\prime\prime}$}

Having reviewed the procedure for deriving $H_{\rm s}^{(k)}$ for
nearest-neighbor hopping $t$ only, we now
discuss how to
include second and third nearest neighbor hopping $t^{\prime}$ and $t^{\prime\prime}$.
We start with an extended
 Hubbard Hamiltonian:
\begin{eqnarray}
H_{\rm H} &= &T + T' + T'' + V\label{HHtt-0} \\ & = &  -t
\sum_{i,j_1;\sigma} c^\dagger_{i,\sigma}c_{j_1,\sigma} -t^{\prime}
\sum_{i,j_2;\sigma}
c^\dagger_{i,\sigma}c_{j_2,\sigma}\nonumber\\
\label{HHtt}\\
& - & t^{\prime\prime} \sum_{i,j_3;\sigma}c^\dagger_{i,\sigma}c_{j_3,\sigma} +
U\sum_i n_{i,\uparrow}n_{i,\downarrow} \nonumber,
\end{eqnarray}
where $t^{\prime}$ is the hopping constant to the $2^{nd}$ nearest
neighbor and $t^{\prime\prime}$ to the $3^{rd}$ nearest neighbor (see Fig.
\ref{hops}); $j_\alpha$ is the $\alpha$ nearest neighbor of $i$.
%
%
As above,
we define the operators $T_{m}$, taking into account
$t^{\prime}$ and $t^{\prime\prime}$:
\begin{eqnarray}
T_1 &= \f{1}{2}&\sum_{i,j;\sigma}(-t_{i,j}) n_{i,\bar{\sigma}}
c^\dagger_{i,\sigma}c_{j,\sigma} h_{j,\bar{\sigma}}, \label{newt1}
\end{eqnarray}
\vv \eqn{\begin{array}{ccc} T_{0}&=&\displaystyle
\f{1}{2}\sum_{i,j;\sigma}(-t_{i,j})
\Big(h_{i,\bar{\sigma}}  c^\dagger_{i,\sigma}c_{j,\sigma} h_{j,\bar{\sigma}}\\
& +& n_{i,\bar{\sigma}}  c^\dagger_{i,\sigma}c_{j,\sigma}
n_{j,\bar{\sigma}}\Big) ,
\end{array}\label{newt0}}
\vv
\begin{eqnarray}
T_{-1}&=&\f{1}{2}\sum_{i,j;\sigma}(-t_{i,j}) h_{i,\bar{\sigma}}
c^\dagger_{i,\sigma}c_{j,\sigma} n_{j,\bar{\sigma}},
\label{newtmoins1}
\end{eqnarray}
where $t_{i,j} \= t$ if $i$ and $j$ are nearest neighbors,
$t_{i,j}=t^{\prime}$ if
$i$ and $j$ are second nearest neighbors,
and $t_{i,j}=t^{\prime\prime}$ if $i$ and $j$ are
third nearest neighbors and $0$ otherwise. The commutation
relations in (\ref{commut}) do not change and the expression for
the unitary transformation in (\ref{unitary}) remains formally the
same. On the other hand, the resulting spin Hamiltonian gets
drastically modified by the inclusion of $t^{\prime}$ or $t^{\prime\prime}$. In
particular, many new {\it plaquette} or ring exchange terms are
generated. The complete expression for the spin Hamiltonian is
reported in Appendix (\ref{ann}). As in previous
work~\cite{mac,Delannoy}, and even more so here, given the much
increased complexity of the Hamiltonian, a computer program was
used to collect together all the terms in the projection of
Eq.~\ref{Tr} which ultimately lead to a globally SU(2) invariant
effective spin-only Hamiltonian $H_{\rm s}^{(4)}(t,t^{\prime},t^{\prime\prime},U)$.

\section{Physical Results}

\label{Results}

\subsection{Spin wave dispersion}

\subsubsection{Spin wave calculation}

\label{SW-calculation}

In order to obtain the magnetic excitation spectra presented in
Fig. \ref{fig_coldea} and Fig. \ref{fig_coldea_fit_JY} we perform
a $1/S$ spin wave calculation. The spin operators are written in
terms of
boson operators through a
Holstein-Primakoff\cite{kitt,Holstein} 1/S expansion for a
bipartite N\'eel ordered square lattice:

\eqn{\begin{tabular}{cc}
Sublattice a  &  Sublattice b \\
$\left\{\begin{tabular}{ccl} $S_i^z$ & $=$& $S-a_i^{\dagger}a_i$
\\  $S_i^+$&$ =$& $\sqrt{2S-a_i^{\dagger}a_i} \, a_i $ \\ $ S_i^-$
&$ =$& $  a_i^{\dag}\sqrt{2S-a_i^{\dagger}a_i}$\end{tabular}\right.$ & $\left\{\begin{tabular}{ccl}
$S_j^z$ & $=$& $-S+b_j^{\dagger}b_j$  \\  $S_j^-$&$ =$&
$\sqrt{2S-b_j^{\dagger}b_j} \, b_j $ \\ $ S_j^+$ &$ =$& $
b_j^{\dag} \sqrt{2S-b_j^{\dagger}b_j}$\end{tabular}\right.$
\end{tabular}
}

In reciprocal space, one obtains to order $S$ the following
general expression of the spin Hamiltonian
$H_{\rm s}^{(4)} \approx H_{{\rm s},0}^{(4)}\openone +H_{\rm s,quad}^{(4)}$,
where $H_{{\rm s},0}^{(4)}$ is the classical ground state energy and
$H_{\rm s,quad}^{(4)}$ is given by
\eqn{H_{\rm s,quad}^{(4)}\=
\sum_{\bf k} A_{\bf k}\left(a_{\bf k}^{\dag}a_{\bf k} \+ b_{\bf
k}^{\dag}b_{\bf k} \right) \+ B_{\bf k}\left(a_{\bf
k}^{\dag}b_{\bf k}^{\dag} \+ a_{\bf k}b_{\bf k} \right).
\label{Hquad} }
Note that here and elsewhere in the paper, the
sums in reciprocal space are over the magnetic Brillouin zone,
which for a square lattice with $N$ spins contains $N/2$ sites.
$H_{\rm s,quad}^{(4)}$ can be diagonalized through a Bogoliubov
transformation
\eqn{\left\{\begin{array}{ccl}
a_{\bf k} & = & u_{\bf k} \alpha_{\bf k} + v_{\bf k} \beta_{\bf k}^\dag\\
b_{\bf k} & = & u_{\bf k} \beta_{\bf k} + v_{\bf k} \alpha_{\bf
k}^\dag
\label{bogolub}
\end{array}\right\},}
giving
\eqn{H_{\rm s,quad}^{(4)} \= \sum_{\bf k} \epsilon_{\bf k} (\alpha_{\bf k}^\dag
\alpha_{\bf k} \+ \beta_{\bf k}^\dag \beta_{\bf k} + 1).}
%
%
%
and where
\begin{eqnarray}
{u_{\bf k}}^2 &=&{ {B_{\bf k}}^2\over{2\epsilon_{\bf k}(A_{\bf k}-\epsilon_{\bf k})}}
\nonumber\\
{v_{\bf k}}^2 &=&{ {A_{\bf k}-\epsilon_{\bf k}}\over{2\epsilon_{\bf k}}}.
\end{eqnarray}
The magnons for a given wave vector ${\bf k}$ are thus two fold
degenerate with eigenfrequencies~\cite{kitt}:
\begin{equation}
\label{sw-energies}
\epsilon_{\bf k} = \sqrt{A_{\bf k}^2 -  B_{\bf k}^2}    \;\;\; .
\end{equation}
In this expression, the functions $A_{\bf{k}}$ and $B_{\bf{k}}$
 take into account all the bilinear and ring exchange
interactions entering $H_{\rm s}^{(4)}$ and
given in Appendix \ref{ann}.
 Detailed expressions
for $A_{\bf{k}}$ and $B_{\bf{k}}$ can be found in Appendix \ref{Asw}.

The staggered magnetization operator, $M_{\rm s}^\dagger$,
is, in reference to sublattice a,
defined conventionally for a spin-only Hamiltonian as
\begin{eqnarray}
M_{\rm s}^\dagger & = & \frac{1}{(N/2)} \sum_{i=1}^{i=N/2} S_i^z .
\end{eqnarray}
Rewriting the previous equation in ${\bf k}$ space and introducing
the Bogoliubov transformation above,
we arrive at a standard expression~\cite{kitt}:
\begin{eqnarray}
M_{\rm s}^\dagger & = & S - \frac{2}{N}  \sum_{\bf k} a_{\bf k}^\dagger a_{\bf k}  =
          S- \frac{2}{N} \sum_k
      (  {u_{\bf k}}^2 \alpha_{\bf k}^\dagger \alpha_{\bf k}
        +{v_{\bf k}}^2 \beta_{\bf k} \beta_{\bf k}^\dagger \nonumber \\
 &+ &   {\rm off-diagonal} \; {\rm terms}      ) \; .
\end{eqnarray}
At zero temperature, all
$\langle \alpha_{\bf k}^\dagger \alpha_{\bf k}\rangle$
 and
$\langle \beta_{\bf k}^\dagger \beta_{\bf k}\rangle$ are zero,
and one has for the ground state expectation value of $M_{\rm s}^\dagger$,
$\langle M_{\rm s}^\dagger \rangle$:
\begin{eqnarray}
\label{zero-point}
\langle M_{\rm s}^\dagger \rangle
& = & S - \frac{2}{N}  \sum_{\bf k} {v_{\bf k}}^2 \nonumber \\
& = & S + \frac{1}{2} - \frac{2}{N}
    \sum_{\bf k} \frac{ A_{\bf k} }{ 2\epsilon_{\bf k}} \; .
 \end{eqnarray}

\subsubsection{Magnon self energy: renormalization factor $Z_c({\bf k})$}

The $1/S$ correction to Eq.~(\ref{zero-point}) vanishes
for the Heisenberg antiferromagnet,
with the next corrections appearing at order $1/S^2$ only \cite{Igarashi}.
This explains why spin wave calculations give
a very good estimate of zero point quantum fluctuations in this
model.
On the other hand, the terms to quartic order in magnon operators in
$H_{\rm s}^{(4)}$ give a
contribution to the diagonal part of
the magnon self energy~\cite{Igarashi}.
This gives rise to a $1/S$ correction to the spin-wave
energies $\epsilon_{\bf k}$ in Eq.~(\ref{sw-energies}).
This renormalization of the magnon
energy scale is important for the quantitative comparison between
our calculation and the experimental results of Coldea {\it et al.}
\cite{coldea01} and for the subsequent determination of the
parameters $\{ t,t^{\prime},t^{\prime\prime},U \}$ and hence we evaluate it for our effective
spin Hamiltonian.

Expanding the spin Hamiltonian to  fourth order in the boson
operators of the Holstein-Primakoff transformation, one finds:
\eqn{H_{\rm s}^{(4)}\= H_0^{(4)} \+ H_{\rm s,quad}^{(4)} \+ H_{\rm
s,quart}^{(4)} \;\;. }
$H_{\rm s,quart}^{(4)}$ is now treated as a first order
perturbation to the quadratic spin-wave Hamiltonian $H_{\rm
s,quad}^{(4)}$. To this order the magnon energy is shifted but the
eigenvectors are unchanged and magnon-magnon interactions are not
generated. The shift is found by diagonalizing the set of $2\times
2$ matrices with elements $\langle 0|\gamma_{\bf k} H_{\rm
s,quart}^{(4)} \gamma_{\bf k}^{'\dagger}|0\rangle$, where
$(\gamma_{\bf k},~\gamma'_{\bf k}) \in \{\alpha_{\bf
k},~\beta_{\bf k}\}$. We hence find a correction $\delta \epsilon_{\bf k}$
to the magnon energy, and a renormalized magnon energy
$\tilde \epsilon({\bf k})$:
\eqn{\epsilon_{\bf k} \longrightarrow \tilde{\epsilon}_{\bf k} \=
\epsilon_{\bf k}+\delta\epsilon_{\bf k} \=
\epsilon_{\bf k} (1 + \Delta_{\bf k})\= \epsilon_{\bf k} Z_c(\bf
k) ,} where we refer to $\Delta_{\bf k}$ as the magnon energy
correction and $Z_c(\bf k)$ the magnon energy renormalization
factor. For the Heisenberg model,
the product $\epsilon_k\Delta_{\bf k}$ corresponds to the
leading contribution to the magnon self energy calculated in
Ref.~[\onlinecite{Igarashi}]. To order $t^2/U$ and with $t^{\prime}$
and $t^{\prime\prime}$ set to zero,
$\Delta_{\bf k}$ is uniform over the whole
Brillouin Zone and we find:
\begin{equation}
Z_c({\bf k}) \,\simeq\, 1.1579,
\label{Igarashi-Zc}
\end{equation}
as obtained in Ref.~[\onlinecite{Igarashi}].
Details of this calculation are given in Appendix \ref{Zc}.

\begin{figure}[h!]
\includegraphics[scale=0.5]{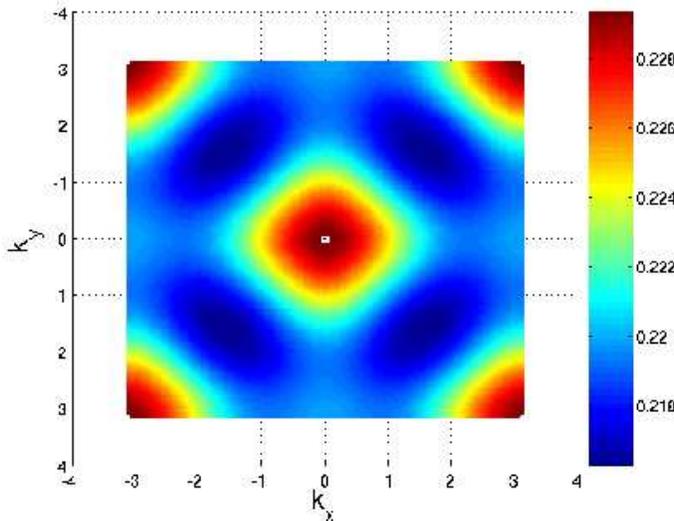}
\caption{ (Color online) ${\bf k}$ dependence of $\Delta_{\bf k}$. Data is shown
over the structural Brillouin zone of the square lattice. As can
be seen, the dispersion has the symmetry of the zone corresponding
to the antiferromagnetic order.
} \label{Zc_fig}
\end{figure}

When non-zero values of $t^{\prime}$ and $t^{\prime\prime}$ are included, $\Delta_{\bf
k}$ is no longer constant over the Brillouin zone, but the
two-fold degeneracy for each wave vector ${\bf k}$ remains. The
renormalizing factors $Z_c(\bf k)$ are calculated numerically. For
a given set of $t^{\prime}$ and $t^{\prime\prime}$ values, a finite size scaling
analysis was used to extrapolate the sums over ${\bf k}'$ in
Appendix \ref{Zc} over the Brillouin zone to the thermodynamic
limit. In Fig. \ref{Zc_fig} we show the dispersion of $Z_c({\bf
k})$ obtained for the parameter set ($t,t^{\prime},t^{\prime\prime},U$) (see
Eq.~\ref{best-fit}) giving the best fit to the magnon dispersion
data~\cite{coldea01} from Coldea {\it et al.}, as described in the
next section. For these values the mean value of $\Delta_{\bf k}$
over the Brillouin zone is found to be
\eqn{ \langle Z_c({\bf k}) \rangle \,\simeq\, 1.219.}
We also calculate the dispersion of $\Delta_k$ over the Brillouin
zone for this set of parameters, finding:
\eqn{\frac{\sigma(\Delta_{\bf k})}{{\langle \Delta_{\bf k} \rangle}_{\bf k}}
\,\simeq\, 0.33 \times 10^{-3}~.}
where $\sigma$ is the standard deviation of the distribution of
values. Details of the calculation can be found in Appendix
\ref{Zc}.

From our results here one can see that the magnon energy
renormalization $Z_c({\bf k})$,
calculated to order $1/S$, is indeed an important
element of the quantitative comparison between our microscopic
theory and the experimental results of Coldea {\it et
al.}~\cite{coldea01}. The values obtained for $Z_c({\bf k})$
change the energy scale of the magnon dispersion by around $20\%$
of what would have been obtained if the a bare $Z_c({\bf k})=1$
had been used.
The $Z_c$ factors
therefore have proportionate consequences for the values of
$U$ and $t$ deduced by comparisons with experiment.

\subsubsection{Experimental data and previous results}

\begin{figure}
\includegraphics[scale=0.5]{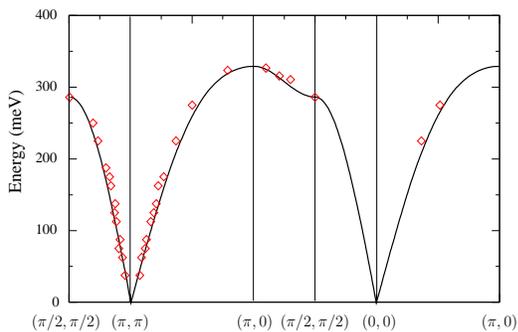}
\caption{(Color online) Magnon energy in La$_2$CuO$_4$, as a function of wave
vector ${\bf k}$, across the Brillouin zone
at a temperature of 10 K. The experimental data
(red squares) and fit (full line) are from reference
\cite{coldea01}. The $\{t,t^{\prime},t^{\prime\prime},U\}$
fitting  parameters are given
in (\ref{Coldea-values}). } \label{fig_coldea}
\end{figure}

The magnon energy spectrum for La$_2$CuO$_4$ obtained through
inelastic neutron scattering~\cite{coldea01} is shown in
Fig.~\ref{fig_coldea}.
The data follow a trajectory through the
Brillouin zone covering, in particular, the line along the
magnetic zone boundary with wave vector in the interval ${\bm
q}_{\rm BZ}\=[(\pi,0),(\pi/2,\pi/2)]$, where the previously
discussed downward dispersion is manifest. Also included in this
figure is the fit of Ref.~[\onlinecite{coldea01}] using the
Hamiltonian~(\ref{SS}). The parameters used there to make the fit
at a temperature of
10 K are~\cite{coldea01} \eqn{
\left\{\begin{array}{lcD{.}{.}{4}l}
t/U & = & 0.135 & \pm \, 0.03             \\
t &= & 0.30     & \pm \, 0.02 \, {\rm eV} \\
U &= & 2.3      & \pm \, 0.4  \, {\rm eV}
\end{array}\right\}.
\label{Coldea-values}
}

While the above fit of Coldea {\it et al.}~\cite{coldea01} is
clearly good, the numerical values should be compared with those
from experiments and band calculations on
other cuprate materials. For example,
values found by fitting ARPES results for
the half-filled compound Sr$_2$CuO$_2$Cl$_2$ containing CuO$_2$
planes are~\cite{Leung:1997,Damascelli,Tohyama:2000}
\eqn{\left\{\begin{array}{lcD{.}{.}{4}l}
t/U & = & 0.1 &\pm 0.05~, \\
t^{\prime}/t & = & -0.35 &\pm 0.05~,~\\
t^{\prime\prime}/t & = & 0.22 &\pm 0.05~,~ \\
t & = & 0.35 & {\rm eV} \\
U &= & 3.5 & {\rm eV}
\end{array}\right\}.\label{exp-values}}
One can see that the energy scale, set by $U$
and the ratio $t/U$ in Eq.~(\ref{exp-values}) are
quite different from that found by Coldea {\it et
al.}~\cite{coldea01}. Also, while the effects of $t^{\prime}$ and $t^{\prime\prime}$
are neglected in Ref.~[\onlinecite{coldea01}],
the ARPES results unambiguously
show that $t^{\prime}$ and $t"$ have non-negligible values within the
CuO$_2$ planes~\cite{Damascelli,Tohyama:2000}, although band
structure calculations\cite{pickett:1989,Andersen,Andersen:2001}
on LaCu$_2$O$_4$ and ARPES experiments on the lightly doped
system~\cite{Yoshida:2005,Ino:1999,Ino:2002,Damascelli} do suggest
somewhat smaller values of $t^{\prime}$ than those
in Eq.~(\ref{exp-values})
for Sr$_2$CuO$_2$Cl$_2$.
In other words, based on various experiments on a number of cuprates,
one would have ``anticipated'' a $t/U$ value for La$_2$CuO$_4$ somewhat
smaller than the value in
Eq.~(\ref{Coldea-values})
from Ref.~[\onlinecite{coldea01}],
although the experimental uncertainty allows for
some overlap between the two.
Most importantly, from a theoretical
point of view, $t/U=0.135$ corresponds to $U=7.4t$, which is
smaller than the tight binding band width of
$8t$.
 This value may therefore be a little small \cite{Vekic:1993},
to insure that the material remains within the Mott insulating
phase. The error bars do allow values up to $U=9.5$ ($t/U=0.105$),
which would push
the model further into the insulating regime. Nevertheless,
it is quite possible that the analysis of the data in
Ref.~(\onlinecite{coldea01}) using the model $H_{\rm s}^{(4)}$ in
Eq.~(\ref {spinHS4}) leads to an underestimation of $U$.

From this discussion
it is clear that a more detailed
analysis of the experimental magnon dispersion
that accounts for $t^{\prime}$ and $t^{\prime\prime}$ is needed.

\begin{figure}
\includegraphics[scale=0.5]{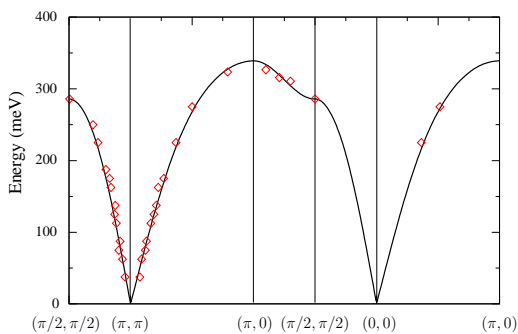}
\caption{(Color online) Magnon energy in La$_2$CuO$_4$, as a function of wave
vector ${\bf k}$, across the Brillouin zone
at a temperature of 10 K. The experimental
data (red squares) are from reference \cite{coldea01}. The fit is
made using the parameters given by
 (\ref{best-fit}). }
\label{fig_coldea_fit_JY}
\end{figure}

\subsubsection{Fitting procedure and fit to the experimental data}

\label{Fitting}

Our theory,
$H_{\rm s}^{(4)}(t,t^{\prime},t^{\prime\prime},U)$ in Appendix A,
now contains four independent parameters, $t,t^{\prime},t^{\prime\prime},U$
which we fit to the ensemble of experimental data points from
Ref.~[\onlinecite{coldea01}].
As we show in Appendix~\ref{Zc}, $Z_c({\bf k})$ is a function of
$t/U,t^{\prime}/t,t^{\prime\prime}/t$ and $\bf k$. It thus should be calculated over
the ensemble of data points for each iteration of the fitting
algorithm. As the expression for $Z_c({\bf k})$ contains
inner sums ($\sum_{\bf k'}$)
 over the Brillouin zone, the fitting procedure is time consuming.
 A finite size scaling analysis of the
convergence of these sums towards the thermodynamic limit
is discussed
in Appendix \ref{Zc}.
The fitting procedure is as follows.
We choose a first point at wave vector
$\bf k$ with experimental magnon energy $\epsilon_{\rm exp}({\bf k})$
and minimize the quantity
$\delta({\bf k}) = |\tilde \epsilon({\bf k})-
\epsilon_{\rm exp}({\bf k})|$ with respect to $t,t^{\prime},t^{\prime\prime},U$. From
this first fit we extract a factor $4t^2/U$ that fixes the energy
scale for the ensemble of points and allows us to write $\tilde
\epsilon({\bf k})$ as $(4t^2/U)$ times a function of
$t/U,t^{\prime}/t,t^{\prime\prime}/t$ and $\bf k$.
This energy scaling step
 can be achieved
for many values of $t/U,t^{\prime}/t$ and $t^{\prime\prime}/t$ and the constraints on
these variables are not very high at this stage. The best fit
parameter set is now found by minimizing the ensemble of variables
$\{\delta({\bf k}) (U/4t^2) \}$ with respect to $t/U,t^{\prime}/U,t^{\prime\prime}/U$,
using least squares fit.  Once the final value of
$t/U$ is established, the value of $t,$ and hence $U$, $t^{\prime}$ and $t"$,
are deduced. The values found are:
\eqn{\left\{\begin{array}{lcD{.}{.}{4}l}
t/U & = & 0.126 &\pm \, 0.03 {\rm } ~, \\
t^{\prime}/t & = & -0.327 &\pm \, 0.05 {\rm } ~,~\\
t^{\prime\prime}/t & = & 0.153 &\pm \,  0.05 {\rm } ~,~ \\
t & = & 0.422 & {\rm eV} \\
U &= & 3.34 & {\rm eV}
\end{array}\right\}.
\label{best-fit}
}
The error bars, obtained through the calculation of the
least-squares function  $\chi^2$, represent a deviation of less
than $5\%$ from the best fit values.

If the dispersion in $Z_c(\bf k)$ is ignored, it only needs to be
calculated once for each iteration of the ensemble of points, thus
considerably speeding up the procedure. Ignoring the dispersion
and taking
$Z_c({\bf k})= \langle Z_c({\bf k}) \rangle ={\rm constant}$,
the best fit parameter set for this case gives the values:

\eqn{\left\{\begin{array}{lcD{.}{.}{4}l}
t/U & = & 0.121 &\pm 0.03~, \\
t^{\prime}/t & = & -0.313 &\pm 0.05~,~\\
t^{\prime\prime}/t & = & 0.167 &\pm 0.05~,~ \\
t &= & 0.430 & {\rm eV}\\
U &= & 3.75 & {\rm eV}
\end{array}\right\}.}

Comparing the two sets, taking the size of the error bars into
account and considering the fact that $Z_c(\bf k)$ represents only
the leading correction to the magnon energies,
we conclude that the dispersion of $\Delta_{\bf k}$
does not have a significant effect
on the determination of the parameters.

The best fit
 to the experimental data using parameter set (\ref{best-fit})
is shown in Fig.~(\ref{fig_coldea_fit_JY}).
Comparing with Fig.~(\ref{fig_coldea}), one observes that
introducing $t^{\prime}$ and $t"$ leads to a clear improvement in the
quality of the fit through the experimental points. One might
have expected
 this, given the increased number of free parameters.
However,
comparing the set (\ref{best-fit}) with (\ref{exp-values}),
 one can
also see that the parameters correspond more realistically with
those thought to describe other cuprates such as, for example,
SrCuO$_2$Cl$_2$. This is particularly the case for the energy
scales $t$ and $U$ which are found to be larger than those in
Eq.~(\ref{Coldea-values}). The ratios $t^{\prime}/t$ and $t"/t$ are also
in fair agreement with values for other cuprates and,
perhaps most interesting, the fitting
procedure even finds the correct signs for these ratios. The ratio
$t/U$ is slightly reduced compared with (\ref{Coldea-values}) and
more in line with a value that one would expect for a system in
the Mott insulating regime, although the change is less
spectacular than in the case of the absolute energy scales $t$ and
$U$, with the difference in the $t/U$ value remaining  within the
experimental error bars. However, considering the overall
comparison between calculations and the experiments of
Ref.~[\onlinecite{coldea01}], it would appear that adding all
terms in $t,t^{\prime}$ and $t^{\prime\prime}$ up to order $1/U^3$ does allow a
quantitative description of the magnon dispersion in
La$_2$CuO$_4$, which is an improvement over the equivalent
procedure which only includes the $t$ and $U$ parameters.

\subsubsection{Comments on the influence of $t^{\prime}$ and $t^{\prime\prime}$}

Referring to Appendix \ref{ann}, one can see that the
inclusion of $t^{\prime}$ and $t^{\prime\prime}$ and all four hop processes involving
$t,t^{\prime}$ and $t^{\prime\prime}$ introduces a large
 number of new ring exchange terms. From this
observation one might ask which of the many terms make the
most difference and could we have got away with a less complex
effective spin Hamiltonian? In this section we address these
questions and justify our choice to do a complete calculation of
all four hop processes.

The first question is whether or not we can limit ourselves to
just one extra parameter $t^{\prime}$ or $t^{\prime\prime}$. This point has already
been discussed in the Introduction (Section \ref{second-n}). We
argued there,  (Eq.~(\ref{J2eff})), that $t^{\prime}$ increases the
strength of the antiferromagnetic second neighbour exchange, $J_2
\rightarrow t^4/U^3 + 4(t^{\prime})^2/U $. This in turn modifies the
effective second neighbour coupling in the spin wave analysis,
$J_2^{\rm eff} \rightarrow 4(t^{\prime})^2/U - 16t^4/U^3$ making it less
strongly ferromagnetic, or even antiferromagnetic if $t^{\prime}$ is
sufficiently large. As the negative magnon dispersion along the
zone boundary~\cite{coldea01} is a consequence of a ferromagnetic
(i.e. negative) $J_2^{\rm eff}$, one expects that fitting the calculated magnon
dispersion to the experimental data will require a larger value of
$t/U$ if $t^{\prime} \ne 0$ but $t^{\prime\prime}$ is ignored. This is indeed the case;
introducing $t^{\prime}$ to order $(t^{\prime})^2/U$ and setting $t^{\prime\prime}=0$, as well
as neglecting all terms of order $(t^{\prime}t)^2/U^3$ and $(t^{\prime})^4/U^3$, the following
best fit parameters are found:

\eqn{\left\{\begin{array}{lcD{.}{.}{4}l}
t/U & = & 0.166&\pm 0.016 \;\;\;                , \\
t^{\prime}/t & = & -0.24 &\pm 0.01 \;\;\;               ,\\
\end{array}\right\}.\label{fit_tp}}
The results are very close to those found from a random phase
approximation~\cite{singh02:_spin_la_cu0,Peres:2002} and also from
quantum Monte Carlo calculations~\cite{Gagne}. It is clear that
the addition of $t^{\prime}$ only has taken the parameter set in the wrong
direction: $t/U$ is now further from most estimates than the case
when $t^{\prime}=0$. Also, with $U=6t$ the system could be very near the
metallic phase~\cite{Vekic:1993},
and not deep in the Mott insulating
phase, as is the case for La$_2$CuO$_4$. As we discussed in the
introduction, for $U=6t$ the $t^{\prime}=0$ model is right at the
metal-insulator transition and the presence of a finite $t^{\prime}$
increases even further the value of $U$ necessary to
position the system
in the insulating phase~\cite{Rousseau:2006,Tremblay:2006}.

The leading effect of third-neighbor hopping $t^{\prime\prime}$ is also to
increase an exchange interaction, this time $J_3$, through the
relation :
\eqn{J_3\= 4\f{t^4}{U^3} \longrightarrow J_3\= 4\f{t^4}{U^3} +
4\f{{(t^{\prime\prime})}^2}{U}.}
However, in this case antiferromagnetic (positive) exchange also
contributes, as does the ring exchange term,
 to a negative (downward) dispersion along the magnetic zone
boundary. Including just $t^{\prime\prime}$, one would therefore expect the
best fit parameters for the data to include a particularly small
ratio of $t/U$, which is indeed the case. Hence we see that, as
far as this characteristic dispersion is concerned, $t^{\prime}$ and $t^{\prime\prime}$
play opposing roles. It is therefore clear that one needs to
include both hopping constants to get a good fit to the
experimental data as we have illustrated in
Fig.~\ref{fig_coldea_fit_JY} with parameters
(Eq.~(\ref{exp-values})) similar to those in other
cuprates~\cite{Kim:1998,Schuttler,Andersen}.

Given that individually,  terms of order $t^2t^{\prime 2}/U^3$ are only
one tenth of the magnitude of terms of order $t^{\prime 2}/U$ or of the terms
of order $t^4/U^3$, one might expect the good fit reported in
Section(~\ref{best-fit}) to be also obtained when taking into
account terms of order $(t^{\prime})^2/U$ and $(t^{\prime\prime})^2/U$ only, and
ignoring the host of more complex four hop terms generated by the
further neighbor hopping. Not so! If one neglects them and only
considers $t^{\prime}$ and $t^{\prime\prime}$ to order $4(t^{\prime})^2/U$ and $4(t^{\prime\prime})^2/U$),
one obtains the following set of best fit parameters:

\eqn{\left\{\begin{array}{lcD{.}{.}{4}l}
t/U & = & 0.14 &\pm 0.03~, \\
t^{\prime}/t & = & 0 &\pm 0.0,~\\
t^{\prime\prime}/t & = & 0 &\pm 0.0~,~ \\
t &= & 0.303 & {\rm eV},\\
U&=& 2.14 & {\rm eV}
\end{array}\right\}.\label{not-best-fit}}

Within numerical error this is the same data set as Coldea {\it et
al.}(~\ref{Coldea-values}), with $t^{\prime}/t=t^{\prime\prime}/t=0$. The rather
surprising and interesting conclusion is therefore that, on their
own, the two-hop further neighbor terms make no net contribution.
Rather, it is the four-hop processes that make the difference. It
seems that their small value is compensated for by their
multiplicity; that is, by the fact that a given site $i$ appears
in a large number of diagrams of order $1/U^3$. Hence, their
contribution is increased by an order of magnitude and in this
intermediate coupling regime, the global effect of the
near-neighbor four hop terms, the majority of which are ring
exchanges, is as important as that for the two hop further
neighbor processes (Note that as $t^{\prime}$ and $t"$ are both zero in
Eq.(\ref{not-best-fit}), $Z_c({\bf k})$ is once again given by
~(\ref{Igarashi-Zc}) over the whole zone). An alternative
procedure would be to impose some partial constraints on the
parameters $(t,t^{\prime},t^{\prime\prime},U)$, using values obtained from, say, band
structure calculations. Results of such fits are briefly discussed
in Appendix ~\ref{Constrained-Fits}.

\begin{figure}[h!]
\centering \subfigure[$\chi^2$ with odd powers of $t^{\prime}/t$ and
$t^{\prime\prime}/t$ excluded \label{fit1}]{
\includegraphics[scale=0.3]{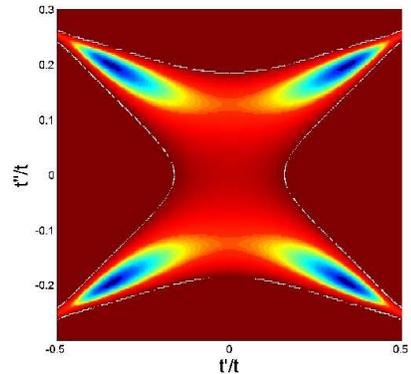}
} \subfigure[$\chi^2$ with odd powers of $t^{\prime}/t$ and $t^{\prime\prime}/t$
included\label{fit2}]{
\includegraphics[scale=0.3]{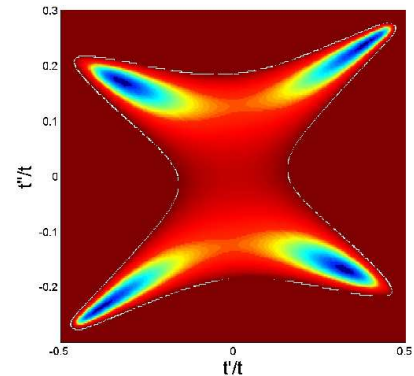}}
\caption{ (Color online)
Goodness of fit parameter $\chi^2$ (grey scale) as a
function $t^{\prime\prime}/t$ and $t^{\prime}/t$ with the terms with odd powers of
$t^{\prime}/t$ and $t^{\prime\prime}/t$ excluded (\ref {fit1}) and included
(\ref{fit2})}\label{fit}
\end{figure}

Finally, we remark that in our fitting procedure, the symmetry is
broken between data sets with  positive and negative values for
the ratio $t^{\prime}/t^{\prime\prime}$. The best fit occurs for a negative ratio
$t^{\prime}/t^{\prime\prime}$ (see Eq.~(\ref{exp-values})). Recall that a canonical
particle-hole transformation at fixed chemical potential changes
the sign of all hopping integrals and leaves the filling
invariant, since $2-n=n$ when $n=1$ \cite{hirsch:1989,Li-Eckern}.
 An additional change of phase of the
creation-annihilation operators on one of the sublattices by $\pi$
restores the sign of nearest-neighbor hopping $t$, but not that of
$t^{\prime}$ and $t^{\prime\prime}$. This shows that $t^{\prime}/t$ and $t^{\prime\prime}/t$ can have
arbitrary sign but that the sign of $t^{\prime}/t^{\prime\prime}$ is physically
relevant. The symmetry breaking between positive and negative
signs for $t^{\prime}/t^{\prime\prime}$ can only be accessed by including four hop
processes. As one can see from Appendix \ref{ann}, most of the
spin interaction terms are even in powers of $t$, $t^{\prime}$ and $t^{\prime\prime}$.
However, a small number of terms have odd powers of $t^{\prime}$ and
$t^{\prime\prime}$. These are the  terms that determine the sign of the hopping
ratios. This can be seen in Fig.~\ref{fit} where we show the
$\chi^2$ parameter from the fitting algorithm as a function of
$t^{\prime}/t$ and $t^{\prime\prime}/t$ (the value of $t/U$ being set to the best fit
value). If the terms proportional to $t^2t^{\prime}t^{\prime\prime}/U^3$ are
deleted from $H_{\rm s}^{(4)}$
(see Fig. \ref{fit1}), the $\chi^2$ parameter has four fold
symmetry and there are four points of best fit independently of
the sign of $t^{\prime}/t$ and $t^{\prime\prime}/t$. On the other hand, when these
terms are included (see Fig.~\ref{fit2}), the symmetry is clearly
broken. The minima for positive ratio $t^{\prime}/t^{\prime\prime}$ become broader and
shallower, while those for negative sign are narrower and deeper.
The fitting procedure hence favors the two points with a negative
sign for the ratio $t^{\prime}/t^{\prime\prime}$. To lift the remaining degeneracy we
have to appeal to band structure
calculations~\cite{pickett:1989,Andersen,Andersen:2001} or ARPES
experiments~\cite{Yoshida:2005,Ino:1999,Ino:2002,Damascelli,Tohyama:2000},
both of which find $t^{\prime}/t$ negative, hence $t^{\prime\prime}/t$ positive.
%


\subsection{Magnetization and spin Hamiltonian}

Having found a set of parameters $(t,t^{\prime},t^{\prime\prime},U)$ that suitably
describes the experimental spin-wave dispersion data, we now
explore how this new set and the derivation of the
effective spin Hamiltonian, $H_{\rm s}^{(4)}(t,t^{\prime},t^{\prime\prime},U)$
affects the value of the zero temperature N\'eel order parameter.

\subsubsection{Magnetization operator}


In the construction of effective theories all operators, ${O}_{\rm H}$,
defined in the original model must be canonically transformed
before they can be exploited
in a calculation within the spin only theory.
 That is, within the effective theory,
${O}_{\rm H}$ becomes ${O}_{\rm s}=e^{i {\cal S}} {O}_{\rm H} e^{-i {\cal S}}$ and
the expectation value in the ground state is defined by
\begin{equation}
\left<{O}\right>\,=\, \frac{_{\rm H}\langle 0|{O}_{\rm H}|0
\rangle_{\rm H}} {_{\rm H}\langle 0|0 \rangle_{\rm H}}
\,=\,\frac{_{\rm s}\langle 0|{O_{\rm s}}|0\rangle_{\rm s}} {_{\rm
s}\langle 0 |0\rangle_{\rm s}} \;\;\; . \label{O2}
\end{equation}

Here $|0\rangle_{\rm H}$ and $|0\rangle_{\rm s}=e^{i{\cal
S}}|0\rangle_{\rm H}$ are the ground state wave vectors in the
original Hubbard (H) and spin-only (S) models. We have recently
shown~\cite{Delannoy} that incorporating this transformation
 has important consequences for the
ground state magnetization as one moves into the intermediate
coupling regime. On application of this procedure on the
magnetization operator we find~\cite{Delannoy} new quantum
fluctuations arising from the charge delocalization over closed
virtual loops of electronic hops. These spin independent
fluctuations allow us to reconcile an apparent paradox concerning
the behavior of the ground state magnetization, as a function of
$t/U$. As discussed above, the leading effect of including
processes to order $t(t/U)^3$ is to introduce effective second and
third neighbor spin interactions (see Eq.~(\ref{reno})).
Within the lowest $1/S$ order
spin wave approximation, the resulting effective ferromagnetic
second neighbor interaction reduces the transverse quantum spin
fluctuations and stabilizes the N\'eel order of the spin model.
This would naively seem to imply
 that the staggered
magnetization should be an
increasing function of $t/U$ as the system departs from the
Heisenberg limit, $t/U\rightarrow 0$; a result which is difficult
to justify on physical grounds.
Indeed, one might expect that as $t/U$ is increased,
the enhanced electron mobility would lead to a
progressive return to the non-magnetic
metallic state, with the staggered moment
decreasing as $t/U$ increases,
as found in Ref.~[\onlinecite{schrieffer}], rather than increasing.
In Ref.~[\onlinecite{Delannoy}]
we showed that this is indeed the case: the new quantum
fluctuations, or ``charge fluctuation'' channel, arising from
virtual doubly occupied states, counter the effect of generating a
ferromagnetic second neighbor interaction. Consequently, the
ground state magnetization for the Hubbard model, when calculated
using the spin-only description, {\it is} indeed
a decreasing function of
$t/U$. The main steps of the calculation are reviewed below and
then extended to include the hopping parameters $t^{\prime}$ and $t^{\prime\prime}$.

Within the
Hubbard Hilbert space, the staggered
magnetization operator $M_{\rm H}^\dagger$ is defined by:

\eqn{M_{\rm H}^{\dag}\,=\,\f{1}{N}\sum_{i}S_i^z e^{-i {\bf Q} \cdot
{{\bf r}_i} },\label{Mag}}
where:
\eqn{ {\bf Q}\,=\,(\pi,\pi)\,\,\,\mbox{and}\,\,\,\,S_i^z
\,=\,\f{1}{2} \sum_{s_1,s_2} c_{i,s_1}^{\dag}
\mathbf{{\sigma}}_{s_1,s_2}^zc_{i,s_2}.}
The magnetization in the ground state $| 0 {\rangle}_{H}$
of the Hubbard model is defined as:
\eqn{M =  {_{\rm H}{\langle}} 0 |M_{\rm H}^{\dag} | 0 {\rangle}_{\rm H}.}
Working within the effective spin theory, we have to express all
the operators in the spin language. Mathematically this means that
the unitary transformation in (\ref{unitary}) has to be applied to
all operators including $M_{H}^{\dag}$ :
\eqn{ M_{\rm s}^{\dag} \=  e^{i {\cal S}} M_{\rm H}^{\dag} e^{-i {\cal
S}}.}

The commutation relations between $M_{H}^{\dag}$ and the different
operators in (\ref{unitary}) are~\footnote{\underline{Proof}
\\
\[[T_1,M_{\rm H}^{\dag}] \,=\, -\f{t}{2N} \sum_{i,j,k,\sigma,\sigma'}
{\left[n_{i\bar{\sigma}} c_{i{\sigma}}^{\dag} c_{j{\sigma}}
h_{j\bar{\sigma}},~\hat{\sigma}^z_{\sigma,\sigma'}n_{k,\sigma'}\right](-1)^k,}\]
\[[T_1,M_{\rm H}^{\dag}] \,=\, -\f{t}{2N} \sum_{i,j,k,\sigma,\sigma'}{n_{i\bar{\sigma}}
\left\{
\left[c_{i{\sigma}}^{\dag},~n_{k,\sigma'}\right]c_{j{\sigma}}
\,+\,c_{i{\sigma}}^{\dag}
\Big[c_{j{\sigma}},~n_{k,\sigma'}\Big]\right\}}\]
\vspace*{-0.5cm}\[\hspace*{5.5cm}\times h_{j\bar{\sigma}}
\hat{\sigma}^z_{\sigma,\sigma'} (-1)^k,\] hence,
\[[T_1,M_{\rm H}^{\dag}] \,=\,-\frac{t}{2N} \sum_{i,j,\sigma}
{n_{i\bar{\sigma}}c_{i{\sigma}}^{\dag}c_{j{\sigma}}h_{j\bar{\sigma}}
\left(-(-1)^{i}\, +\,(-1^j\right))
\hat{\sigma}^z_{\sigma,\sigma}.}\]
\noindent
Since
$(-1)^i\,=\,-(-1)^j$ for nearest neighbors, we get the final
result and can generalize it for $T_{-1}$ and $T_0$.}
%
\eqn{[T_1,M_{\rm H}^{\dag}]\,\hat{=}\,\f{1}{N}\tilde{T_1}\,=\,\f{t}{N}
\sum_{i,j,\sigma} n_{i,\bar{\sigma}} c_{i,\sigma}^{\dag}
c_{j,\sigma} h_{i,\bar{\sigma}} (-1)^i
\mathbf{\hat{\sigma}}_{\sigma,\sigma}^z,\label{commut1}}
\eqn{[T_{-1},M_{\rm H}^{\dag} ]\,\hat{=}\,\f{1}{N}\tilde{T}_{-1}\,=\,
\f{t}{N} \sum_{i,j,\sigma} h_{i,\bar{\sigma}} c_{i,\sigma}^{\dag}
c_{j,\sigma} n_{i,\bar{\sigma}} (-1)^i
\mathbf{\hat{\sigma}}_{\sigma,\sigma}^z,\label{commut2}}
\begin{eqnarray}
[T_0,M_{\rm H}^{\dag}] & \hat{=} &\f{1}{N}\tilde{T_0}\,=\, \f{t}{N}
\sum_{i,j,\sigma} \left(n_{i,\bar{\sigma}} c_{i,\sigma}^{\dag}
c_{j,\sigma} n_{i,\bar{\sigma}}
(-1)^i \mathbf{\hat{\sigma}}_{\sigma,\sigma}^z \right. \nonumber\\
 & +& \left.  h_{i,\bar{\sigma}} c_{i,\sigma}^{\dag}
c_{j,\sigma} h_{i,\bar{\sigma}} (-1)^i
\mathbf{\hat{\sigma}}_{\sigma,\sigma}^z\right).\label{commut3}\end{eqnarray}
Expressed in terms of these operators, the
staggered magnetization operator reads\cite{Delannoy}:
\eqn{M_{\rm s}^{\dag}\,=\, M \+ \f{1}{U} \left(\tilde{T}_1
-\tilde{T}_{-1}\right)\+ \f{1}{2U^2}\left(\tilde{T}_{-1}T_1 -
T_{-1}\tilde{T}_{1}\right)\label{effmag}.}
The symmetry associated with half filling ensures that
contributions to this expression corresponding to odd numbers of
hops,
such as the second term in the right-hand side of Eq.~(\ref{effmag}),
 are zero. Writing the expression in terms of $S=1/2$ spin
operators\cite{Delannoy} we find:
%
\begin{eqnarray}
 M_{\rm s}^{\dag} & = &  \f{1}{N}\left(\sum_i S_i^z e^{i{\bf Q}\cdot {{{\bf r}_i}}} -
 \f{2t^2}{U^2} \sum_{<i,j>}\left\{ S_i^z \,-\, S_j^z\right\}
{\rm e}^{i{\bf Q}\cdot {{{\bf r}_i}}}\right). \nonumber \\
\label{one}
\end{eqnarray}
In the case of periodic boundaries, or in the thermodynamic limit
where boundaries can be neglected, all sites become equivalent and
the correction to the magnetization operator becomes a
multiplicative factor\cite{Delannoy}:
\eqn{M_{\rm s}^{\dag} \= \tilde{M}_{\rm s}^{\dag} \left(1 - 8\f{t^2}{U^2}
\right),\label{plop}}
where $\tilde{M}_{\rm s}^{\dag}\=\f{1}{N}\sum_i S_i^ze^{i {\bf Q}
\cdot {{{\bf r}_i}}}$ is the naive expression for the
magnetization operator in the spin language\cite{Delannoy}.

\subsubsection{Role of $t^{\prime}$ and $t^{\prime\prime}$}

Inclusion of the higher order hopping constants $t^{\prime}$ and $t^{\prime\prime}$
introduces extra paths for charge delocalization and we might
expect new charge fluctuation terms to appear.
The hopping operators $T_m$ in Eqs.~
(\ref{newt1},\ref{newt0},\ref{newtmoins1}), can be generalized to
include $t^{\prime}$ and $t^{\prime\prime}$ and the commutation relations in
Eqs.~(\ref{commut1},\ref{commut2},\ref{commut3}) are redefined:
\eqn{[T_m,M_{\rm H}^{\dag}]\,\hat{=}\,\f{1}{N}\tilde{T}_m,}
\eqn{[T'_m,M_{\rm H}^{\dag}]\,\hat{=}\,\f{1}{N}\tilde{T'}_m,\label{tildpr}}
\eqn{[T''_m,M_{\rm H}^{\dag}]\,\hat{=}\,\f{1}{N}\tilde{T''}_m,\label{tildsec}}
for hops involving
 $t$,$t^{\prime}$ and $t^{\prime\prime}$, respectively. The staggered
magnetization operator (\ref{effmag}) is thus also generalized.
Dropping the first order terms in $1/U$ that do not
contribute to the expectation value in a singly-occupied state,
$M_{\rm s}^\dagger$ takes the form:
\eqn{\begin{array}{ccl}
M_{\rm s}^{\dag} &= & M \+ \f{1}{2U^2}\left\{\left(\tilde{T}_{-1} +\tilde{T'}_{-1} + \tilde{T''}_{-1} \right)\left(T_1+T'_1+T'_1\right)\right. \\
\\
&-& \left.\left(T_{-1}+T'_{-1}+T''_{-1}\right)\left(\tilde{T_{1}}+
\tilde{T'_{1}} + \tilde{T''_{1}}\right)\right\}.
\end{array}
\label{effmagtt}}
However, it turns out that the operators $\tilde{T}'_m$ and
$\tilde{T}''_m$ defined in Eqs. ~(\ref{tildpr},\ref{tildsec}) are
identically equal to $\hat{\mathbf 0}$
~\footnote{\underline{Proof}:\\Same calculation than in
Eq.~(\ref{commut1},\ref{commut2},\ref{commut3}), but for second or
third nearest neighbors, $(-1)^i\,=\,(-1)^j$.}.
From this result,
it follows that the first corrections to Eq.~(\ref{plop})
that depend on $t^{\prime}$ and $t^{\prime\prime}$ appear beyond the second order in the
perturbation scheme.
%
%
Hence we find that
to order $(t^{\mu}/U)^2$ the expression for $M_{\rm
s}^{\dag}(t,t^{\prime},t^{\prime\prime})$ is unchanged and given by Eq. ~(\ref{plop}).
However, this does not mean that introducing $t^{\prime}$ and $t^{\prime\prime}$ has no
effect on the expectation value for the staggered magnetization. Firstly,
the inclusion of $t^{\prime}$ and $t^{\prime\prime}$ in the magnon dispersion modifies
the ensemble of parameters $(t,t^{\prime},t^{\prime\prime},U)$ that best fit the data.
Secondly, the zero point fluctuations are controlled by
$(t/U,t^{\prime}/t,t^{\prime\prime}/t)$,
through the ${\bf k}$ dependence of
$u_{\bf k}$, $v_{\bf k}$, $A_{\bf k}$ and $B_{\bf k}$
in Section ~\ref{SW-calculation} and this, as we  discuss below,
directly affects $\tilde M_{\rm s}^\dagger$.

The charge renormalization factor has also been calculated using a
Hartree-Fock mean-field method as in Ref.~[\onlinecite{schrieffer}], in
which transverse spin fluctuations are neglected (see Appendix B).
This calculation
on a square lattice gives the same factor, $1-8(t/U)^2$,
independently of $t^{\prime}$ and $t^{\prime\prime}$. Physically this comes about
because,
in the spin-density wave ground state (i.e. N\'eel order,
${\bf Q}=(\pi,\pi)$),
electron hops between two sites that belong
to the same N\'eel ordered sublattice
are prohibited by the Pauli principle.
Hence, same-sublattice hops by $t^{\prime}$ and $t^{\prime\prime}$ do not
contribute to charge renormalization to order $(t^{\prime})^2/U$, $(t^{\prime\prime})^2/U$.

\subsubsection{Consequences}

An experimental estimate of the zero temperature sublattice
ordered moment in La$_2$CuO$_4$ has been determined from neutron
scattering. Lee {\it et al.}\cite{Lee} find
\eqn{M_{\rm experiment} \= g \moy{M_{\rm s}^{\dag}} \mu_B \=  0.55 \pm
0.05 \mu_B~, \label{M}}
where $g$ is the Cu$^{2+}$ Land\'e $g$-factor.
The $g$-factor for Cu$^{2+}$ in layered cuprates such
as La$_2$CuO$_4$ has apparently
not been determined experimentally~\cite{Kivelson}.
However,
 a typical value
for a distorted octahedral environment is  given by Abragam and
Bleaney~\cite{Abragam}, ${g\simeq 2.2}$. We use this value to
compare theoretical estimates of the ordered moment with the
experimental value. A spin wave calculation for the
nearest-neighbor Heisenberg antiferromagnet
gives~\cite{Reger:1988,Gross,Runge} $\moy{M_{\rm s}^{\dag}}^{\rm
nn} \simeq 0.304$,  which is in good agreement with the value
obtained from quantum Monte Carlo.
 This gives for the moment,
$M^{\rm sw}=g\langle M_{\rm sw}^\dagger\rangle^{\rm nn}$, $M^{\rm
sw} \approx 0.67 \mu_B $, well in excess of the above $0.55 \mu_B$
value determined experimentally~\cite{Lee}. Using the parameter set of Coldea
{\it et al.}~\cite{coldea01} but neglecting charge renormalization
leads to an {\it increased} expectation value:
$\moy{M_{\rm s}^{\dag}}\simeq 0.32$, taking the moment
 to $0.70 \mu_B$,
hence further from the experimental estimate
of Ref.~[\onlinecite{Lee}].
Including
the charge renormalization factor~\cite{Delannoy}  $1-8t^2/U^2$
of  Eq.~(\ref{plop})
gives $\moy{M_{\rm s}^{\dag}}\simeq 0.27$ and $M^{(t)}\simeq 0.59\mu_B$ in
better agreement with experiment.
Including both the charge
renormalization and the effects of $t^{\prime}$ and $t^{\prime\prime}$ we find
\eqn{\moy{M_{\rm s}^{\dag}}^{(tt^{\prime}t^{\prime\prime})} \simeq 0.235,}
from which we obtain
\eqn{
\moy{M^{\dag}}^{(tt^{\prime}t^{\prime\prime})}
=g  \moy{M_{\rm s}^{\dag}}^{(tt^{\prime}t^{\prime\prime})} \simeq 0.52 \mu_B \;\;\;  ,}
which is in even better agreement with experiment. Given the
uncertainty in the experimental measurement value
in Eq.~(\ref{exp-values}) and in the theoretical
value that should be
taken for $g$, it is difficult to make in-depth comment on the difference
between $\langle M\rangle^{tt^{\prime}t^{\prime\prime}}$ and $\langle M\rangle ^{t}$.
However, the results do serve to
illustrate the importance of the charge renormalization at this
level of approximation. Here, it is an essential element in the
quantitative agreement with experiment, and it is only through the
inclusion of  spin independent quantum corrections
(e.g. virtual double site-occupancy)
that quantitative agreement can be achieved.

\section{Discussion}

It is known from ARPES measurements and band structure
calculations that $t^{\prime}$ and $t^{\prime\prime}$ are not negligible compared to
$t$ in a number of copper oxide materials. We have thus introduced
second and third neighbor hopping parameters, $t^{\prime}$ and $t^{\prime\prime}$, into
the Hubbard Hamiltonian and included all closed four hop virtual
electronic pathways into the canonical transformation to derive an
effective spin-only Hamiltonian for the case of a half filled
band. This Hamiltonian contains many ring exchange terms. These
kinds of terms have been the subject of much discussion
lately~\cite{Singh-ring,Lauchli,Motrunich,Hermele,Pujol-ice,Delannoy}.
Through this calculation we are able to test the capacity of the
one band Hubbard model to describe the magnetic properties of the
antiferromagnetic parent high-temperature superconductor
La$_2$CuO$_4$. We find in general good quantitative agreement
between the predictions of the $t-t^{\prime}-t^{\prime\prime}-U$ model and experimental
measurements of the magnetic properties of the
compound~\cite{Note:3-band}.

An exact solution of the spin-only Hamiltonian we have derived is
clearly far out of reach. In fact, because of the frustration and
ensuing sign problem introduced by the various frustrating
bilinear spin-spin couplings and the many ring exchange terms,
even a quantum Monte Carlo attack on the low temperature
properties of this model seems difficult to envisage in the near
future. However, it has long been known that  a spin-wave analysis
up to leading order in a $1/S$ expansion accurately reproduces the
zero temperature staggered moment in the Heisenberg model on a
square lattice~\cite{Reger:1988}. This is because of the
cancellation of the magnon-magnon interaction terms to second
order in $1/S$ for this particular lattice~\cite{Igarashi}. This
suggested that a similar spin wave analysis can be a good starting
point to describe the observed magnetic excitation spectrum in
La$_2$CuO$_4$ and this is indeed what we have found in this paper.

In our spin wave calculations, we calculated the magnetization to
leading order in $1/S$ and included the first order correction to the
classical spin wave frequencies.
The magnon
energies are thus renormalized by a spin-wave renormalization
factor $Z_c({\bf k})$. Since the classical spin-wave frequencies
are already of order $1/U^3$, we only retained terms of order
$t^2/U$, $(t^{\prime})^2/U$ and $(t^{\prime\prime})^2/U$ in calculating $Z_c({\bf k})$.
Hence, to this order, it does not depend on the ring exchange
terms which are of order $1/U^3$. In the end, $Z_c({\bf k})$
raises the energy scale by  about $20\%$ over the whole of the
Brillouin zone, allowing good agreement between experiment and
theory without the introduction of any arbitrary scale factors.

To make contact with experiments on La$_2$CuO$_4$ the spin wave
calculation was set up starting from the classical
antiferromagnetic N\'eel ground state. Within this framework, the
general effect of the ring exchange is to reduce the transverse
quantum spin fluctuations rather than increase them (see however
the discussion of charge renormalization in the following
paragraph).
With the parameters coming out of the calculation, we therefore
seem far from entering an exotic quantum  phase
driven from large quantum spin fluctuations about
the N\'eel ordered state. Yet,
the ring exchanges are found to play a very important
role. In particular, it seems that the large numbers of such terms
and their large dimensionless prefactors (see Appendix A)
outweighs their small O($1/U^3$) amplitude, although a more extensive study
is required here to understand this point in detail. Further, ring
exchange plays a crucial role in determining the relative sign of
the ratios $t^{\prime}/t$ and $t^{\prime\prime}/t$. There is experimental evidence from
ARPES measurements and band structure calculations that $t^{\prime}/t$ is
negative while $t^{\prime\prime}/t$ is positive in a number of cuprates. In our
calculation there are terms of order $t^2t^{\prime}t^{\prime\prime}/U^3$ which break
the $t^{\prime\prime}/t^{\prime} \rightarrow -t^{\prime\prime}/t^{\prime}$ symmetry
 in the effective spin-only Hamiltonian $H_s^{(4)}$,
allowing us to deduce the correct sign for the ratio of $t^{\prime}/t^{\prime\prime}$
compared with experiment. In our procedure, we determined
unconstrained values of $U,t,t^{\prime}$ and $t^{\prime\prime}$ by comparing our magnon
dispersion data with that from experiments on
La$_2$CuO$_4$~\cite{coldea01}. The best fit is found with one
positive ($t^{\prime\prime}/t$) and one negative ($t^{\prime\prime}/t^{\prime}$) ratio, exactly as
in  ARPES experiments and band calculations and, for example, our
best fit data set is in good agreement with those from the ARPES
experiments on Sr$_2$CuO$_2$Cl$_2$ (compare Eq.(\ref{best-fit})
with Eq.(\ref{exp-values})).

At the level of approximation at which the canonical
transformation is performed~\cite{Harris,mac,Delannoy}, finite
charge mobility renormalizes the magnetic moment calculated from
the spin only Hamiltonian by a factor
$(1-8t^2/U)$ ~\cite{Delannoy}. We found that $t^{\prime}$ and $t^{\prime\prime}$ do not
contribute further to this factor. This is because direct second
and third neighbor hops take electrons from one site to another on
the same sublattice, hence virtual double occupancy is excluded
by the Pauli exclusion principle. The expectation value for the
zero temperature staggered moment coming from the spin only
Hamiltonian is thus scaled by the same $(1-8t^2/U)$ factor as in
the case where $t^{\prime}=t^{\prime\prime}=0$ ~\cite{Delannoy}. The relevance of this
term can be estimated by comparing theoretical and experimental
estimates of the total magnetic moment. Here, we also find good
agreement between the estimates of Lee {\it et al} in
Ref.~[\onlinecite{Lee}] and our calculated value. The charge
renormalization is important here, as it scales the magnetization
 obtained by considering solely transverse spin fluctuations,
 by an amount clearly in excess of the
experimental error bars. The comparison with experiment therefore
provides an implicit illustration of its importance in the real
material.

Our results also bear on the question of the appropriate ratio of the bandwidth $\simeq 8t$ to the interaction strength $U$. First, note that on purely theoretical grounds, when $U$ is
of the order of the tight-binding band width, one
expects the Hubbard model to undergo an insulator to metal
transition, resulting in
a breakdown of the perturbation expansion in $t/U$ ~\cite{Hartmann:2002,Rousseau:2006}. In our calculations, since the next order
terms in the expansion of $t/U$ is smaller than the leading term,
there is no evidence of such a breakdown. In fact, for the value $t/U \simeq 0.126$ (or $U \simeq 8t$) that we found, we argue that the Hubbard model is definitely in the Mott insulator regime, as is necessary for the expansion to be valid. Indeed, at $t\prime=t^{\prime\prime}=0$, one can extract from quantum Monte Carlo calculations \cite{Vekic:1993} that the Mott gap should close around $t/U \simeq 0.167$ (or $U \simeq 6t$), a value consistent with recent estimates from quantum cluster calculations~\cite{Park-cond-matt, Gull:2008}. Although the critical $t/U$ in general depends on $t\prime$,~\cite{Rousseau:2006,Kyung:2006} for $t\prime=-0.3t$ on the square lattice, the critical $t/U$ is still very close to the value appropriate for $t\prime=0$~\cite{Kyung:2008}. One can also check that for $t/U \simeq 0.126$, the single-particle spectral weight displays bands associated with antiferromagnetic excitations that disperse with $J$ and are distinct from Hubbard bands further away from the Fermi energy.~\cite{KyungPseudogap:2006} When the gap is induced purely  antiferromagnetic fluctuations (Slater mechanism), the distinct Hubbard bands are absent.~\cite{Vilk:1997}. In addition, for insulating behavior induced by antiferromagnetic fluctuations, the potential energy decreases when insulating behavior occurs.~\cite{Kyung:2003a} This is not obseved for $t/U \simeq 0.126$.~\cite{Paiva:2001}.

For $t\prime=t^{\prime\prime}=0$, the best fit
value of $t/U \simeq 0.14$ found by Coldea {\it et al.}~\cite{coldea01} places La$_2$CuO$_4$ dangerously close
to the insulator to metal transition discussed above. However, with the inclusion of
$t^{\prime}$ and $t^{\prime\prime}$ and of all ring exchange terms to order $1/U^3$, we
find a ratio $t/U \simeq 0.126$ which is decreased compared with the initial fit
of Coldea {\it et al.}, which did not include these further neighbor hoppings.  This places the
ratio $t/U$ for La$_2$CuO$_4$ within the Mott insulating
phase of the Hubbard model discussed in the previous paragraph, by contrast with the result found in Ref.~\onlinecite{Comanac}.
If we had obtained the opposite result,
namely that the best fit for $t/U$ increases
in the presence of $t^{\prime}$ and $t^{\prime\prime}$,
the whole approach would have become extremely
doubtful as a description of La$_2$CuO$_4$.
Another approach, starting for example from
the three-band model~\cite{Hartmann:2002},
would have become necessary. Instead, the result that we find here
for $t/U$ gives a consistent picture where the parameters of the
 one-band Hubbard model describe an insulator at half-filling.
These parameters are also in agreement with those describing the
 doped insulator \cite{Andersen:2001}, including the sign and
 magnitude of the ratio $t^{\prime}/t^{\prime\prime}$. In fact, a non-vanishing $t^{\prime\prime}$
is necessary to obtain a consistent picture since, without it, $t/U$
would increase~\cite{Gagne} compared with its $t^{\prime}=t^{\prime\prime}=0$ value,
an undesirable state of affairs, as argued above.

To close, and reiterate,
one of the noteworthy results of this work is that the presence of
the ring exchange terms in the effective spin-only theory allow us
to determine the relative sign of $t^{\prime}$ and $t^{\prime\prime}$. More generally,
the theory presented herein, restricted to the reduced Hilbert
space of spin degrees of freedom only, gives results for the {\it
electronic band parameters} that are compatible with ARPES
experiments and theoretical band calculations for a variety of
cuprate materials. In addition, the value of $U \simeq 8t$ that we find is consistent with the origin of the insulating behavior in parent high-temperature superconductors being mostly due to Mott Physics. As a final word, we propose that our results
provide further evidence that the $t-t^{\prime}-t^{\prime\prime}-U$ one-band
Hubbard model in the intermediate coupling regime gives a
consistent unified description of high-temperature cuprate
superconductors and of their parent insulating phases.





\section{Acknowledgements}

We thank A. Chernyshev,
R. Coldea, A. del Maestro, F. Delduc, M. Ghaznavi, A.
La\"uchli, C. Lhuilier, D. McMorrow, D. Poilblanc, M. Roger, N.
Shannon, G. Sawatzky, and F. Vernay for useful discussions.
In particular, we are grateful to
R. Coldea for providing us with the inelastic neutron data on
La$_2$CuO$_4$. A.-M.S.T. acknoledges discussions with A.J. Millis at the Aspen Center for Physics. Partial support for this work was provided by the
NSERC of Canada and the Canada Research Chair Program (Tier I)
(M.G. and A.T.), Research Corporation and the Province of Ontario
(M.G.), FQRNT Qu\'ebec (A.T.) and a Canada$-$France travel grant
from the French Embassy in Canada (M.G. and P.H.). M.G. and A.T.
acknowledge support from the Canadian Institute for Advanced
Research. M.G. thanks the University of Canterbury (UC) for
financial support and the
hospitality of the Department of Physics and Astronomy at UC
where part of this work was completed.

\appendix


\section{Spin Hamiltonian with $t$, $t^{\prime}$ and $t^{\prime\prime}$ \label{ann}}

The purpose of this appendix is to give the new spin-spin
interaction terms appearing in the spin-only Hamiltonian including
$t^{\prime}$ and $t^{\prime\prime}$ up to order $1/U^3$ in the canonical
transformation. The sole effect of introducing $t^{\prime}$ and $t^{\prime\prime}$ to
order $(t^{\prime})^2/U$ and $(t^{\prime\prime})^2/U$ is to renormalize the coupling
constants already present in the spin Hamiltonian (Eq.
\ref{spinHS4}) generated from the $t-U$ Hubbard model, up to order
$t^4/U^3$. Including $t^{\prime}$ and $t^{\prime\prime}$ up to order, $1/U^3$ generates
many more terms that further renormalize these coupling constants.
However, a number of terms with new ``topologies'' appear and
these play a key role on the results
discussed in this paper. For
reasons of compactness we introduce the following notation:
\eqn{
\begin{array}{ccl}
{\rm P}_1^{i,j,k,l} &=& \left\{(\vect{S_i}\,\cdot\,\vect{S_j})\,(\vect{S_k}\,\cdot\,\vect{S_l})\right.\\
&&\left. +(\vect{S_i}\,\cdot\,\vect{S_l})\,(\vect{S_k}\,\cdot\,\vect{S_j}) \, -\,(\vect{S_i}\,\cdot\,\vect{S_k})\,(\vect{S_j}\,\cdot\,\vect{S_l})\right\}\\
{\rm P}_2^{i,j,k,l} &=&  \left\{ \vect{S_i}\cdot \vect{S_j}\,+\,\vect{S_i}\cdot \vect{S_k}\,+\,\vect{S_i}\cdot \vect{S_l}\right.\\
&+&\left.\vect{S_j}\cdot \vect{S_k}\,+\,\vect{S_j}\cdot \vect{S_l} +\, \vect{S_k}\cdot \vect{S_l}\,     \right\}
\end{array}
}
\begin{figure}[h!]
\includegraphics[scale = 1]{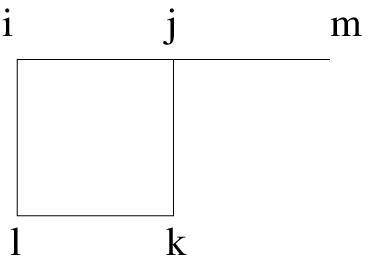}
\caption{Label of the different sites involved in the many-spin terms induced by
$t^{\prime}$ and $t^{\prime\prime}$ to order $1/U^3$}\label{fig2}
\end{figure}
Figure~\ref{fig2} illustrates the sites involved in $t^{\prime}$ and $t^{\prime\prime}$
hopping (see also Fig.~(\ref{hops})).
The symbol \includegraphics[scale=0.5]{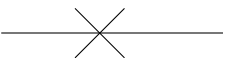} means that
the corresponding sites participate in the expression of the
coupling interaction, whereas the symbol
\includegraphics[scale=0.5]{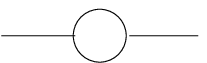} means that these sites are
transparent in the electronic process. That is  the electron
`hops over' the site in question. As there is at present considerable
interest in diluted Mott-Hubbard systems, such as in the
La$_2$Cu$_x$Zn$_{1-x}$O$_4$/
La$_2$Cu$_x$Mg$_{1-x}$O$_4$/
 \cite{Sandvik,Vajk,Castro,Delannoy2},
we keep track, in the derivation of the spin Hamiltonian, of the
occupation of the sites visited by the electrons, so that
$\epsilon_i=1$ if a site $i$ is occupied by a spin and
$\epsilon_i=0$ if the site is not occupied. Here we only consider
the case of $100\%$ Cu$^{2+}$ site occupation ($x=1$).
\begin{center}
\begin{tabular}{cl}
\scalebox{0.4}[0.4]{\includegraphics{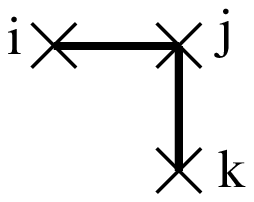}} &\hspace{1cm}  $4 \,\f{{t^{\prime}}^{2}}{U}
\,{{\vec S}_i}\cdot {{\vec S}_k} \left\{ \epsilon_i \epsilon_k\right\}$\\
\scalebox{0.4}[0.4]{\includegraphics{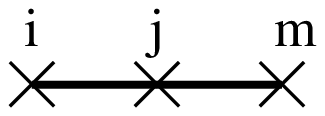}} &\hspace{1cm}  $4 \,\f{{t^{\prime\prime}}^{2}}{U}
\,{{\vec S}_i}\cdot {{\vec S}_m} \left\{ \epsilon_i \epsilon_m\right\}$\\
\scalebox{0.4}[0.4]{\includegraphics{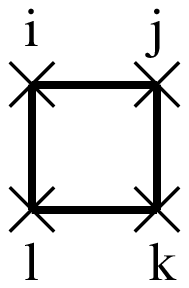}} &\hspace{1cm}  $\,-8 \, \f{{t^{\prime}}^{2}t^{2}}{U^3} \,
\left\{ \epsilon_i \epsilon_j \epsilon_k \epsilon_l  \right\} P_2^{i,j,k,l}$\\
\scalebox{0.4}[0.4]{\includegraphics{Nequal4_1}} &\hspace{1cm}
$4\, \f{{t^{\prime}}^{2}t^{2}}{U^3}\, \left\{ \epsilon_i \epsilon_j
\epsilon_k   \right\} ({{\vec S}_i}\cdot {{\vec S}_j} + {{\vec
S}_j}\cdot
\vec{S_k} )$ \\
\scalebox{0.4}[0.4]{\includegraphics{Nequal4_1}} &\hspace{1cm}
$\,160 \, \f{{t^{\prime}}^{2}t^{2}}{U^3}\, ({{\vec S}_i}\cdot {{\vec
S}_k}) ( {{\vec S}_j}\cdot {{\vec S}_l})\, \left\{ \epsilon_i
\epsilon_j \epsilon_k \epsilon_l \right\}$
\end{tabular}
\end{center}

\begin{center}
\begin{tabular}{cl}
\scalebox{0.4}[0.4]{\includegraphics{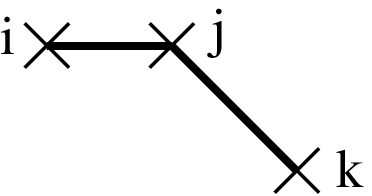}} &\hspace{1cm} $4 \,\f{ {t^{\prime} }^{2}t^{2}}{U^3} \, {{\vec S}_i}\cdot {{\vec S}_k}\, \left\{\epsilon_i \epsilon_j \epsilon_k \right\}$\\
\scalebox{0.4}[0.4]{\includegraphics{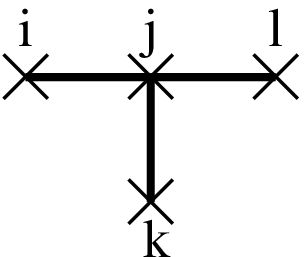}} &\hspace{1cm} $80\f{ {t^{\prime}}^{2}t^{2}}{U^3} \,\epsilon_i \epsilon_j \epsilon_k \epsilon_l \,\,{\rm P}_1^{i,j,k,l}$\\

\end{tabular}
\begin{tabular}{cl}

\scalebox{0.4}[0.4]{\includegraphics{Nequal4_tprime_weared}} &\hspace{1cm} $4\f{ {t^{\prime}}^{2}t^{2}}{U^3} \,\epsilon_i \epsilon_j \epsilon_k \epsilon_l\, (-1)\, \,{\rm P}_2^{i,j,k,l} $ \\
\scalebox{0.4}[0.4]{\includegraphics{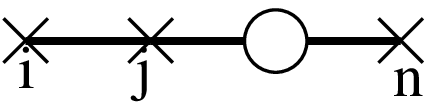}} &\hspace{1cm} $4 \,\f{ {t^{\prime\prime}}^{2}t^{2}}{U^3} \, {{\vec S}_i}\cdot {{\vec S}_n}\, \left\{\epsilon_i \epsilon_j \epsilon_n \right\}$\\
\\
\scalebox{0.4}[0.4]{\includegraphics{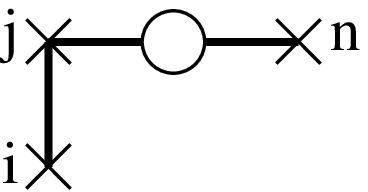}} &\hspace{1cm} $4 \,\f{ {t^{\prime\prime}}^{2}t^{2}}{U^3} \, {{\vec S}_i}\cdot {{\vec S}_n}\, \left\{\epsilon_i \epsilon_j \epsilon_n \right\}$\\
\end{tabular}
\begin{tabular}{cl}

\scalebox{0.4}[0.4]{\includegraphics{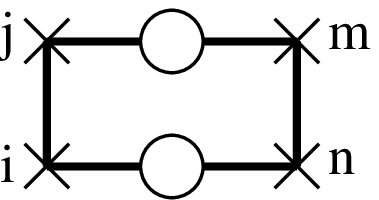}} &\hspace{1cm} $80 \, \f{{t^{\prime\prime}}^{2}t^{2}}{U^3} \,\epsilon_i \epsilon_j \epsilon_m \epsilon_n \,{\rm P}_1^{i,j,m,n}      $ \\
\scalebox{0.4}[0.4]{\includegraphics{Nequal4_tsecond}} &\hspace{1cm} $4 \,\f{ {t^{\prime\prime}}^{2}t^{2}}{U^3} \,\epsilon_i \epsilon_j \epsilon_m \epsilon_n \,(-1)\, \,{\rm P}_2^{i,j,m,n}$ \\
\end{tabular}
\begin{tabular}{cl}
\scalebox{0.4}[0.4]{\includegraphics{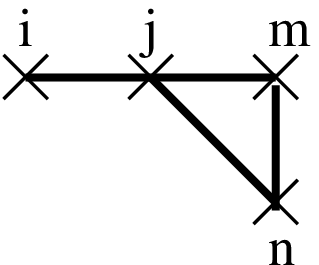}} & \hspace{1cm} $80 \f{{t}^{2}}{U^3} t^{\prime} t^{\prime\prime} \,\epsilon_i \epsilon_j \epsilon_m \epsilon_n \,\,{\rm P}_1^{i,j,n,m} $\\
\scalebox{0.4}[0.4]{\includegraphics{Nequal4_tprimetsecond_weared}} & \hspace{1cm} $4 \f{{t}^{2}}{U^3} t^{\prime} t^{\prime\prime} \, \,\epsilon_i \epsilon_j \epsilon_m \epsilon_n \,(-1)\,\, {\rm P}_2^{i,j,n,m}$\\
\end{tabular}
\begin{tabular}{cl}
\scalebox{0.4}[0.4]{\includegraphics{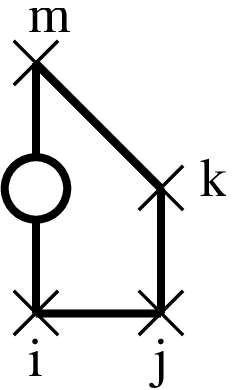}} &  \hspace{1cm} $80\f{ {t}^{2} t^{\prime} t^{\prime\prime}}{U^3} \,\epsilon_i \epsilon_j \epsilon_m \epsilon_k \,\,{\rm P}_1^{i,j,k,m}  $
\end{tabular}

\begin{tabular}{cl}
\scalebox{0.4}[0.4]{\includegraphics{Nequal4_tprimetsecond_biseau}} &  \hspace{1cm} $4 \f{{t}^{2} t^{\prime} t^{\prime\prime}}{U^3} \, \,\epsilon_i \epsilon_j \epsilon_m \epsilon_k \,(-1)\,\,{\rm P}_2^{i,j,k,m} $
\end{tabular}

\begin{tabular}{cl}
\scalebox{0.4}[0.4]{\includegraphics{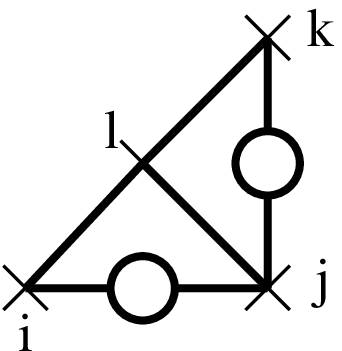}} & \hspace{1cm} $80 \f{{t^{\prime}}^{2} {t^{\prime\prime}}^2}{U^3} \,\epsilon_i \epsilon_j \epsilon_l \epsilon_k \,\,{\rm P}_1^{i,j,k,l}      $\\
\scalebox{0.4}[0.4]{\includegraphics{Nequal4_tprimetsecond_bigtriangle}} & \hspace{1cm}  $4 \,\f{ {t^{\prime}}^{2} {t^{\prime\prime}}^2}{U^3} \,\epsilon_i \epsilon_j \epsilon_l \epsilon_k \,(-1)\,\,{\rm P}_2^{i,j,k,l} $\\
\scalebox{0.4}[0.4]{\includegraphics{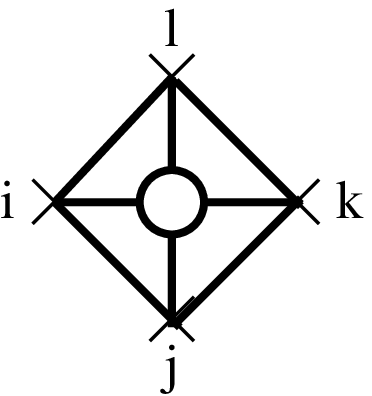}} & \hspace{1cm} $160\f{{t^{\prime}}^{2} {t^{\prime\prime}}^2}{U^3} \,\epsilon_i \epsilon_j \epsilon_l \epsilon_k \left\{({{\vec S}_i}\cdot {{\vec S}_k}) ( {{\vec S}_j}\cdot {{\vec S}_l}) \right\} $\\
& \hspace*{1.cm} $   +  80 \f{{t^{\prime}}^{4}}{U^3} \,\epsilon_i \epsilon_j \epsilon_l \epsilon_k \,\,{\rm P}_1^{i,j,k,l}  $\\
\scalebox{0.4}[0.4]{\includegraphics{Nequal4_tprimetsecond_bigcarre}} & \hspace{1cm} $4 \,\f{{t^{\prime}}^{2} {t^{\prime\prime}}^2}{U^3} \,\epsilon_i \epsilon_j \epsilon_l \epsilon_k \,(-1)\, {\rm P}_2^{i,j,k,l}$
\end{tabular}
\begin{tabular}{cl}
\scalebox{0.4}[0.4]{\includegraphics{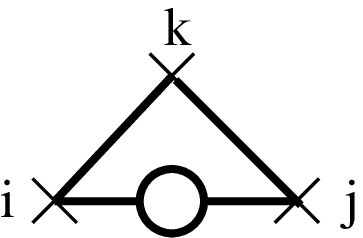}} & \hspace{1cm} $4 \f{{t^{\prime}}^{2} {t^{\prime\prime}}^2}{U^3} \,\epsilon_i \epsilon_j  \epsilon_k \left\{\vect{S_i}\cdot \vect{S_k} + \vect{S_j}\cdot \vect{S_l} \right\}$\\
\scalebox{0.4}[0.4]{\includegraphics{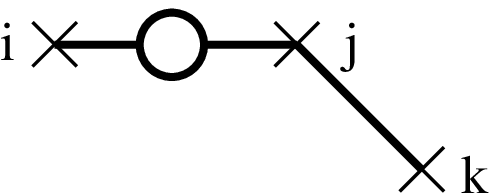}} & \hspace{1cm} $ 4\f{{t^{\prime}}^{2} {t^{\prime\prime}}^2}{U^3} \,\epsilon_i \epsilon_j  \epsilon_k \left\{\vect{S_i}\cdot \vect{S_k} \right\}       $\\
\scalebox{0.4}[0.4]{\includegraphics{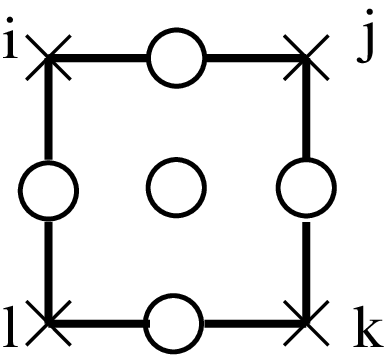}} &  \hspace{1cm} $ 80\f{{t^{\prime\prime}}^{4}}{U^3} \,\epsilon_i \epsilon_j  \epsilon_k  \epsilon_l\,\,{\rm P}_1^{i,j,k,l}    $\\
\scalebox{0.4}[0.4]{\includegraphics{Nequal4_tsecond_4}} &  \hspace{1cm} $ 4\f{{t^{\prime\prime}}^{4}}{U^3} \,\epsilon_i \epsilon_j  \epsilon_k  \epsilon_l\,\,{\rm P}_2^{i,j,k,l}     $\\
\scalebox{0.4}[0.4]{\includegraphics{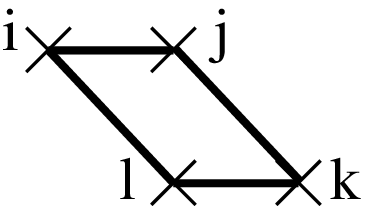}} &  \hspace{1cm} $ 80\f{{t}^{2}{t^{\prime}}^{2}}{U^3} \,\epsilon_i \epsilon_j  \epsilon_k \epsilon_l \,\,{\rm P}_1^{i,j,k,l}    $\\
\scalebox{0.4}[0.4]{\includegraphics{Nequal4_tprimet_paralle}} &  \hspace{1cm} $ 4\f{{t}^{2}{t^{\prime}}^{2}}{U^3} \,\epsilon_i \epsilon_j  \epsilon_k \epsilon_l \,\,{\rm P}_2^{i,j,k,l}     $
\end{tabular}
\end{center}



\section{Mean-field solution of spin Hamiltonian with $t$, $t^{\prime}$ and $t^{\prime\prime}$ \label{An3}}

The purpose of this appendix is  to determine the sublattice
magnetization at zero temperature for the $t-t^{\prime}-t^{\prime\prime}-U$ Hubbard
model using the Hartree-Fock (mean-field) method of
Ref.~[\onlinecite{schrieffer}]. From Eqs.~(\ref{HHtt-0},\ref{HHtt}),
the $t-t^{\prime}-t^{\prime\prime}-U$ Hubbard model is:
\begin{eqnarray}
H_{\rm H} &= &T + T' + T'' + V\nonumber  \\ \\ & = &  -t \sum_{i,j_1;\sigma}c^\dagger_{i,\sigma}c_{j_1,\sigma} -t^{\prime} \sum_{i,j_2;\sigma}c^\dagger_{i,\sigma}c_{j_2,\sigma} -t^{\prime} \sum_{i,j_3;\sigma}c^\dagger_{i,\sigma}c_{j_2,\sigma} \nonumber \\ & &\+ U \sum_i n_{i,\uparrow}n_{i,\downarrow},
\end{eqnarray}
where $j_1$, $j_2$ and $j_3$ are respectively the first, second
and third nearest neighbors of $i$. Fourier transforming this
expression leads to:
\eqn{\begin{array}{ccl}
H_{\rm H} &=&\displaystyle \sum_{{\bf k},\sigma}(\epsilon_{\bf k} + \epsilon_{\bf k}'
+ \epsilon_{\bf k}'' )
c_{k,\sigma}^\dag c_{k,\sigma} \\&+&\displaystyle \f{U}{2N} \sum_{{\bf k},{\bf k'},
{\bf q}}\sum_{\sigma,\sigma',\beta,\beta'} \delta_{ \sigma,\sigma'}
\delta_{\beta,\beta'} c_{{\bf k'},\sigma}^\dag c_{{\bf -k'+q},\beta'}^\dag c_{{\bf -k+q},\beta}c_{{\bf k},\sigma},
\end{array}\label{C1}}
where:
\eqn{\left\{\begin{array}{ccl}
\epsilon_{\bf k} &=& -2t (\cos(k_x) + \cos(k_y)),\\
\\
\epsilon_{\bf k}' &=& -2t^{\prime} (\cos(k_x+k_y) + \cos({k_x-k_y})),\\
\\
\epsilon_{\bf k}'' &= & -2t^{\prime\prime} (\cos(2k_x) + \cos(2k_y)).
\end{array}\right.}

The sublattice magnetization for nesting wave vector ${\bf Q}$ is
defined by:
\eqn{M \= \langle \Omega | S_{{\bf Q}}^z | \Omega \rangle,}
where we consider the case where the magnetization is polarized
along the $\hat{z}$ direction. $\ket{\Omega}$ is the spin density
wave ground state and ${\bf Q}$. The spin density operator $S_{\bf
q}^i$ for arbitrary {\bf q} wave vector is defined by:
\eqn{S_{\bf q}^i \= \f{1}{N} \sum_{{\bf k},\alpha,\beta}
c_{{\bf k+q},\alpha}^\dag \hat{\sigma}^i_{\alpha,\beta} c_{{\bf k},\beta},}
$N$ being the number of sites.

We define new operators $\gamma_{{\bf k},\alpha}^c$ and
$\gamma_{{\bf k},\alpha}^v$ through the Bogoliubov transformation:
\eqn{\left\{\begin{array}{ccl} \gamma_{{\bf k},\alpha}^c & = &
\displaystyle u_{\bf k} c_{{\bf k},\alpha}
+ v_{\bf k} \sum_\beta \hat{\sigma}^3_{\alpha,\beta} c_{{\bf k+Q},\beta},\\
\\
\gamma_{{\bf k},\alpha}^v & = & \displaystyle v_{\bf k} c_{{\bf k},\alpha}
- u_{\bf k} \sum_\beta \hat{\sigma}^3_{\alpha,\beta} c_{{\bf k+Q},\beta}.
\end{array}\right\},\label{C2}}
and inject (\ref{C2}) into the Hartree-Fock
factorization~\cite{schrieffer} of Hamiltonian (\ref{C1}),
finding:
\eqn{E_{\bf k} \= \epsilon_{\bf k}' +  \epsilon_{\bf k}'' \pm \sqrt{\Delta^2 + {\epsilon_{\bf k}}^2},}
where the density wave gap $\Delta$ is given by:
\eqn{\Delta \= -\f{U M}{2}.\label{self-1}}
The Bogoliubov coefficients for this transformation $u_{\bf k}$
and $v_{\bf k}$ are:
\eqn{\left\{\begin{array}{ccl}
{u_{\bf k}}^2 &=& \f{1}{2}\left(1+\f{\epsilon_{\bf k}}{E_{\bf k}-\epsilon_{\bf k}'-\epsilon_{\bf k}'' }\right),\\
{v_{\bf k}}^2 &=& \f{1}{2}\left(1-\f{\epsilon_{\bf k}}{E_{\bf k}-\epsilon_{\bf k}'-\epsilon_{\bf k}'' }\right),\\
\end{array}\right.}
With
the valence band filled and the conduction band empty, since the calculation here is done at zero
temperature, it follows from the above relationships that
\eqn{M \= \langle \Omega | S_{{\bf Q}}^z | \Omega \rangle \=
\f{2}{N}\sum_{\bf k} u_{\bf k} v_{\bf k}~;}
hence:
\begin{equation}
\label{self-2}
 M \= - \f{4}{N}\sum_{\bf k} \f{\Delta}{E_{\bf
k}-\epsilon_{\bf k}'-\epsilon_{\bf k}''}.
\end{equation}
Putting (\ref{self-1}) into (\ref{self-2}) gives the following
self-consistent equation for the sublattice magnetization:
\eqn{\f{1}{U}\= \f{1}{N}\sum_{\bf k} \f{1}{\sqrt{{\epsilon_{\bf
k}}^2 + \left(\f{U M}{2}\right)^2 }}. \label{imp} }
We remark that equation (\ref{imp}) is independent of
$\epsilon_{\bf k}'$ or $\epsilon_{\bf k}''$. Considering
$t/U\ll1$, we can expand the square roots of (\ref{imp}) in
$\epsilon_{\bf k}/U$, after which we find:
\eqn{\f{1}{N}{\sum_{\bf k}}\f{1}{({\epsilon_{\bf k}}^2 +
\Delta^2)^{1/2}} \= \f{1}{N}{\sum_{\bf k}}
\f{2}{UM}\left(1-\f{2{\epsilon_{\bf k}}^2}{U^2
M^2}\right)\=\f{1}{U}.}
%
Further simplifications lead to:
\eqn{M\= 1 \- \f{4}{NM^2} {\sum_{\bf k}} \f{{\epsilon_{\bf
k}}^2}{U^2}.}
To first order in $t/U$ we obtain:
\eqn{M \approx 1 \- \f{4}{N} {\sum_{\bf k}} \f{{\epsilon_{\bf
k}}^2}{U^2}.}
Since ${1\over{N}}{\sum_k} \f{\epsilon_k^2}{U^2} \= \f{4t^2}{U^2}
\f{1}{N} {\sum_k} (\cos(k_x)\+\cos(k_y))^2$, and  \eqn{\f{1}{N}
{\sum_k} (\cos(k_x)\+\cos(k_y))^2\=\f{1}{2}\label{summmm},} we

conclude that:
\eqn{M\= 1 \- 8\f{t^2}{U^2}.\label{plop1}}
This result is equivalent to that obtained through the unitary
transformation method in Eq. (\ref{plop}) if transverse spin
fluctuations are neglected, as they are in the Hartree-Fock
approach. Mathematically, it is the property
$\epsilon_{\bf{k}}=-\epsilon_{\bf{k+Q}}$
that makes nearest-neighbor hopping different from the other two
hops which both obey, instead,
$\epsilon_{\bf{k}}'=\epsilon_{\bf{k+Q}}'$,
$\epsilon_{\bf{k}}''=\epsilon_{\bf{k+Q}}''$. Note that
$\epsilon_{\bf {k}},\epsilon'_{\bf {k}},\epsilon''_{\bf {k}}$ in
this appendix are the quasi-particle energies, which should not be
confused with the spin-wave (magnon) energies in Section
\ref{SW-calculation}.

Physically, since $t^{\prime}$ and $t^{\prime\prime}$ are hops within the same
sublattice (parallel spins) of the classical N\'eel antiferromagnetic solution, in
the strong coupling limit the Pauli principle prohibits hopping
between these sites. Hence, to leading order, $t^{\prime}$ and $t^{\prime\prime}$ cannot change double
occupancy or, in other words, produce charge fluctuations (or ``charge
renormalization'').


\section{Spin Wave Results}\label{Asw}

By introducing the transformations of Eq.~(\ref{bogolub}) in
$H_{\rm s,quad}^{(4)}(t,t^{\prime},t^{\prime\prime},U)$ (equation (\ref{Hquad})),
passing into Fourier space, and collecting all terms, we obtain,
after some tedious but straightforward algebra, the following
expressions for the functions $A_{\bf k}$ and $B_{\bf k}$:
\begin{widetext}
\eqn{\begin{array}{cl}
{\f{U}{(4t^2)}A_{\bf k}}\,= &  4J_1S \+  J_2 S  \Bigg[4\cos(k_x)\cos(k_y)-4\Bigg]
\+ J_3  S  \Bigg[ 2\cos(2k_x) \+ 2\cos(2k_y) - 4 \Bigg]
\\& - J_c  S^3  \Bigg[ 4 \+ 4  \cos(k_x)\cos(k_y) \Bigg] \\
& \+ 40  S^3  \left(\f{t}{U}\right)^2  \left(\f{t^{\prime}}{t}\right)^2
\Bigg[-4 \+ 2\cos(k_x\+k_y)\+2\cos(k_x-k_y)\Bigg]
 \+ 16  S  \left(\f{t}{U}\right)^2  \left(\f{t^{\prime}}{t}\right)^2 \\ &
\+  20 S^3 \left(\f{t}{U}\right)^2  \left(\f{t^{\prime}}{t}\right)^2
 \Bigg[ 16 - 16\cos(k_x)\cos(k_y)\+ 4\cos(2k_x) \+ 4\cos(2k_y)\Bigg]
\\& - S \left(\f{t}{U}\right)^2 \left(\f{t^{\prime}}{t}\right)^2
\Bigg[8(\cos(k_x)\cos(k_y)-1)\+ 4(\cos(2k_x)\+\cos(2k_y))
 \+ 16  \+ 8(\cos(k_x)\cos(k_y)-1)\Bigg] \\&\+
12\,S  \left(\f{t}{U}\right)^2  \left(\f{t^{\prime}}{t}\right)^2  \\
& \+ 20 S^3 \left(\f{t}{U}\right)^2 \left(\f{t^{\prime\prime}}{t}\right)^2
 \Bigg[-16 \+ 8\cos(2k_x) \+ 8\cos(2k_y)\Bigg] \\
& -  S  \left(\f{t}{U}\right)^2 \left(\f{t^{\prime\prime}}{t}\right)^2
\Bigg[32 \+8(\cos(2k_x)\+\cos(2k_y)-2)\Bigg] \\
& - 20 S^3   \left(\f{t}{U}\right)^2  \left(\f{t^{\prime}}{t}\right)
 \left(\f{t^{\prime\prime}}{t}\right)  \Bigg[32 - 8(\cos(2k_x) \+ \cos(k_x-k_y) \+ \cos(k_x\+k_y) \+ \cos(2k_y))\Bigg]
 \\ 
%
& - \left(\f{t}{U}\right)^2  \left(\f{t^{\prime}}{t}\right)  \left(\f{t^{\prime\prime}}{t}\right)  S  \Bigg[32 \+ 16(\cos(k_x)\cos(k_y)) \+  8(\cos(2k_x)\+\cos(2k_y))\Bigg]\\
%
& \+  20  \left(\f{t}{U}\right)^2  \left(\f{t^{\prime}}{t}\right)  \left(\f{t^{\prime\prime}}{t}\right)  S^3  \Bigg[16 - (16 (\cos(k_x)\cos(k_y)-1)) - 8(\cos(2k_x) \+ \cos(2k_y) -2)) \\
&~~~~~~~~ \+ 4(4\cos(k_x)\cos(k_y)-4)  \Bigg]\\
& - 2\left(\f{t}{U}\right)^2  \left(\f{t^{\prime}}{t}\right)  \left(\f{t^{\prime\prime}}{t}\right) S  \Bigg[8  \cos(k_x)\cos(k_y) \+  4\cos(2k_x) \+ 4\cos(2k_y)\Bigg] \\
%
&\+  20 S^3 \left(\f{t}{U}\right)^2  \left(\f{t^{\prime}}{t}\right)^2  \left(\f{t^{\prime\prime}}{t}\right)^2
 \Bigg[-32 \+ 16   (\cos(2k_x) \+ \cos(2k_y)) \+ 16  \cos(k_x)\cos(k_y) \\
&~~~~~~~~-   8\cos(2k_x\+2k_y) - 8\cos(2k_x-2k_y)\Bigg]    \\
& \+ 2 S \left
(\f{t}{U}\right)^2  \left(\f{t^{\prime}}{t}\right)^2  \left(\f{t^{\prime\prime}}{t}\right)^2
  \Bigg[-48 \+ 8  (\cos(2k_x) \+  \cos(2k_y)) \+ 24  (\cos(k_x)\cos(k_y)) \\
&~~~~~~~~\+ 4\left(\cos(2k_x\+2k_y)\+  \cos(2k_x-2k_y)\right)\Bigg] \\
%
&\+  S \left(\f{t}{U}\right)^2  \left(\f{t^{\prime}}{t}\right)^2  \left(\f{t^{\prime\prime}}{t}\right)^2   \Bigg[32\cos(k_x)\cos(k_y) - 32\Bigg] \\
& \+ S \left(\f{t}{U}\right)^2  \left(\f{t^{\prime}}{t}\right)^2  \left(\f{t^{\prime\prime}}{t}\right)^2   \Bigg[-16 \+ 4  (\cos(3k_x\+k_y)  \+\cos(3k_x-k_y) \+ \cos(k_x\+3k_y) \+ \cos(k_x-3k_y) )\Bigg] \\
%
& \+ 20 S^3 \left(\f{t}{U}\right)^2  \left(\f{t^{\prime\prime}}{t}\right)^4
 \Bigg[-4 \+ 4(\cos(2k_x)\+  \cos(2k_y)) - 2\cos(2k_x\+2k_y) - 2\cos(2k_x\+2k_y)\Bigg] \\
& \+ S \left(\f{t}{U}\right)^2  \left(\f{t^{\prime}}{t}\right)^4  \Bigg[(4\cos(2k_x) \+ 4\cos(2k_y)) - 12  \+ 2\cos((2k_x\+2k_y) \+ \cos(2k_x\+2k_y))\Bigg] \\
&\+ 40 S^3 \left(\f{t^{\prime}}{t}\right)^2  \left(\f{t}{U}\right)^2   \Bigg[16(\cos(k_x)\cos(k_y) -1)\Bigg] \\
&\+ 2 S \left(\f{t^{\prime}}{t}\right)^2  \left(\f{t}{U}\right)^2  \Bigg[16\+16\cos(k_x)\cos(k_y)\Bigg]
\end{array}}

\eqn{\begin{array}{cl}
{\f{U}{(4t^2)}B_{\bf k}}\,= & 2  J_1  S  \Bigg[\cos(k_x) + \cos(k_y)\Bigg]- 4 J_c  S^3  \Bigg[\cos(k_x) + \cos(k_y) \Bigg]\\
& + S  \left(\f{t}{U}\right)^2  \left(\f{t^{\prime}}{t}\right)^2  4 \Bigg[\cos(2k_x+k_y)+\cos(2k_y+k_x)  + \cos(2k_x - k_y) + \cos(2k_y-k_x)\Bigg] \\
&+ 80  S^3  \left(\f{t}{U}\right)^2  \left(\f{t^{\prime}}{t}\right)^2  \Bigg[\cos(k_x) + \cos(k_y)\Bigg] \\
& - S  \left(\f{t}{U}\right)^2  \left(\f{t^{\prime}}{t}\right)^2  \Bigg[4(\cos(k_x)+\cos(k_y))+4( \cos(k_x)+\cos(k_y)) + 4(\cos(k_x)+\cos(k_y))\Bigg] \\
& + S  \left(\f{t}{U}\right)^2  \left(\f{t^{\prime\prime}}{t}\right)^2 \Bigg[2(\cos(3k_x)+\cos(3k_y)) +  2\cos(2k_x+k_y) + 2\cos(2k_x-k_y) \\
&~~~~~~~~+ 2\cos(k_x+2k_y) + 2\cos(k_x-2k_y)\Bigg] \\
&+ 20 S^3 \left(\f{t}{U}\right)^2  \left(\f{t^{\prime\prime}}{t}\right)^2   \Bigg[-8(\cos(k_x)+\cos(k_y)) \\

& ~~~~~~~~+ 4 (\cos(2k_x+k_y)+ \cos(2k_x-k_y)+\cos(k_x+2k_y)+\cos(k_x-2k_y))\Bigg]  \\
%
&- S  \left(\f{t}{U}\right)^2  \left(\f{t^{\prime\prime}}{t}\right)^2  \Bigg[ 8(\cos(k_x)+\cos(k_y))+ 4 (\cos(2k_x+k_y)+ \cos(2k_x-k_y)\\
&~~~~~~~~+\cos(k_x+2k_y)+\cos(k_x-2k_y))\Bigg] \\
&- 20 S^3 \left(\f{t}{U}\right)^2  \left(\f{t^{\prime}}{t}\right)  \left(\f{t^{\prime\prime}}{t}\right)  \Bigg[4 (2\cos(k_x)+2\cos(k_y))- 4(\cos(2k_x+k_y) + \cos(2k_x-k_y) \\
&~~~~~~~~+ \cos(2k_y+k_x) +\cos(2k_y-k_x))\Bigg] \\
&- S \left(\f{t}{U}\right)^2  \left(\f{t^{\prime}}{t}\right)  \left(\f{t^{\prime\prime}}{t}\right)  \Bigg[ 24  (\cos(k_x)+\cos(k_y)) + 4 (\cos(2k_x+k_y) + \cos(2k_x-k_y) \\
&~~~~~~~~+ \cos(2k_y+k_x)  + \cos(2k_y-k_x))\Bigg] \\
&+ 20 S^3  \left(\f{t}{U}\right)^2  \left(\f{t^{\prime}}{t}\right)  \left(\f{t^{\prime\prime}}{t}\right)   \Bigg[8(\cos(k_x)+\cos(k_y)) + 8(\cos(k_x)+\cos(k_y)) -4(\cos(2k_x+k_y)\\
&~~~~~~~~+\cos(2k_x-k_y)+ \cos(k_x+2k_y)+\cos(k_x-2k_y))  \Bigg] \\
&- 2 S \left(\f{t}{U}\right)^2  \left(\f{t^{\prime}}{t}\right)  \left(\f{t^{\prime\prime}}{t}\right)    \Bigg[ 4\cos(k_x) +4\cos(k_y) +  2(\cos(2k_x+k_y)+\cos(2k_x-k_y)\\
&~~~~~~~~+\cos(k_x-2k_y)+\cos(k_x+2k_y))\Bigg] \\
%
&+ 40 S^3 \left(\f{t}{U}\right)^2  \left(\f{t^{\prime}}{t}\right)^2    \Bigg[-8(\cos(k_x)+\cos(k_y)) + (2\cos(2k_x-k_y)+2\cos(2k_x+k_y) + 2\cos(k_x+2k_y) \\&~~~~~~~~+  2\cos(k_x-2k_y)) + 4(\cos(k_x)+\cos(k_y))\Bigg] \\
&+ 2 S \left(\f{t}{U}\right)^2  \left(\f{t^{\prime}}{t}\right)^2    \Bigg[12(\cos(k_x)+\cos(k_y))+ (2\cos(2k_x-k_y)+2\cos(2k_x+k_y) + 2\cos(k_x+2k_y) \\
&~~~~~~~~+ 2\cos(k_x-2k_y))\Bigg]
\end{array}}

\end{widetext}


\section{Spin wave renormalization factor $Z_c$}\label{Zc}


We consider the terms in the transformed spin only Hamiltonian of
order $1/S^2$ as a perturbation to the lowest order spin wave
Hamiltonian. Within this framework the perturbation gives quartic
terms in boson creation and annihilation operators, for which we
have to diagonalize the set of $2\times 2$ matrices of elements
$\langle 0|\gamma_{\bf k} H_4 \gamma'^{\dag}_{\bf k}|0\rangle$,
where $(\gamma_{\bf k},~\gamma'_{\bf k}) \in \{\alpha_{\bf
k},~\beta_{\bf k}\}$.

 We first review the situation when $t^{\prime}$ and $t^{\prime\prime}$ are set
equal to zero:
\eqn{\left\{\begin{array}{ccl}
H_2(t) & = &\displaystyle  S J_1(t) \sum_{<i,j>} (a_i^\dag a_i + b_j^\dag b_j) + 2(a_i b_j + a_i^\dag b_j^{\dag}),\\
H_4(t) & = &\displaystyle  -\f{J_1(t)}{2} \sum_{<i,j>} \Big(
a_i^\dag a_i a_i b_j + a_i b_j^\dag b_j b_j +  a_i^\dag a_i^\dag
a_i b_j^\dag \\ && \displaystyle + a_i^\dag b_j^\dag b_j^\dag b_j
\Big) + 2 a_i^\dag a_i b_j^\dag b_j .
\end{array}\right.}
Here, we work to leading order in $t/U$, for which $J_1(t)=
4t^2/U$. Note that, within this appendix, we use a compact
notation: $H_{\rm s,quad}^{(4)}\rightarrow H_2$ and $H_{\rm
s,quart}^{(4)}\rightarrow H_4$. It turns out that the $2\times 2$
perturbation matrix is proportional to the identity matrix. We
find:
\eqn{\begin{array}{ccl}
\delta \epsilon_{\bf k}
&=& \displaystyle \f{1}{2N}J_1(t) \Bigg[\\
&&\displaystyle-2(u_{\bf k}^2 + v_{\bf k}^2)\sum_{\bf k'}  \Big(u_{\bf k'}v_{\bf k'}\big(2\cos(k'_x)+2\cos(k'_y)\big)\Big)\\
&&\displaystyle -\Big(u_{\bf k}v_{\bf k}\big(2\cos(k_x)+2\cos(k_y)\big)\Big)\sum_{\bf k'}2v_{\bf k'}^2 \\
&&\displaystyle + 4(u_{\bf k}^2 + v_{\bf k}^2) \sum_{\bf k'}2v_{\bf k'}^2\\
&&\displaystyle + 4u_{\bf k}v_{\bf k}\sum_{\bf k'} u_{\bf
k'}v_{\bf k'}\big(2\cos(k_x - k'_x)+2\cos(k_y k'_y)\big)\Bigg],
\end{array}\label{eqt}}
from which the magnon energy
correction $\Delta_{\bf k}\equiv \Delta\epsilon_{\bf k}/\epsilon_{\bf k}$ can
be computed.
$\Delta_{\bf k}$ converges numerically to give
\eqn{\Delta_{\bf k} \,\simeq\, 0.1579,}
uniformly over the Brillouin zone, as found previously in Ref.~[\onlinecite{Igarashi}].

When $t^{\prime}$ or $t^{\prime\prime}$ are brought into the picture the leading
effects are to create second and third nearest neighbor
interactions $J_2(t^{\prime})=4(t^{\prime})^2/U$ and $J_3(t^{\prime\prime})=4(t^{\prime\prime})^2/U$
producing new contributions to the quartic Hamiltonian:
\eqn{\left\{\begin{array}{ccl}
H_4(t^{\prime}) &= & \displaystyle J_2(t^{\prime}) \sum_{<<i,j>>}  a_i^\dag a_i a_j^\dag a_j \- \f{1}{4} \left[a_i^\dag a_i a_i a_j^\dag   \right.\\
&&\displaystyle \left.+ a_i^\dag a_j^\dag a_j^\dag a_j  +a_i^\dag a_i^\dag a_i a_j + a_i^\dag a_j^\dag a_j a_j\right]\\
H_4(t^{\prime\prime}) &= & \displaystyle J_3(t^{\prime\prime}) \sum_{<<<i,j>>>} a_i^\dag a_i a_j^\dag a_j \- \f{1}{4} \left[a_i^\dag a_i a_i a_j^\dag \right.\\ &&\displaystyle \left. + a_i^\dag a_j^\dag a_j^\dag a_j +a_i^\dag a_i^\dag a_i a_j + a_i^\dag a_j^\dag a_j a_j\right]
\end{array}\right. }
The contribution coming from $t^{\prime}$  reads:
\eqn{\begin{array}{ccl}

\delta\epsilon_{\bf k}'
 &=& \displaystyle \f{1}{2N} J_2(t^{\prime}) \left[ 2(u_{\bf k}^2 + v_{\bf k}^2) \sum_{\bf k'} 4 v_{\bf k'}^2 \right.\\
&&+ \displaystyle  (u_{\bf k}^2 + v_{\bf k}^2) \sum_{ \bf k'} 2 (u_{\bf k'}^2 + v_{\bf k'}^2) \times \\
&&\hspace*{1cm}\displaystyle \Big(2\cos((k_x+k_y)-(k'_x+k'_y))\,+\, \\&&\hspace*{1cm}\displaystyle 2\cos((k_x-k_y)-(k'_x-k'_y))\Big) \\
&&+\displaystyle     (u_{\bf k}^2 + v_{\bf k}^2) \sum_{ \bf k'} (u_{\bf k'}^2 + v_{\bf k'}^2)\times\\
&&\hspace*{1cm}\displaystyle 2\Big(\cos(k'_x+k'_y)+\cos(k'_x-k'_y)\Big)\\
&&\displaystyle +2(\cos(k_x+k_y)+\cos(k_x-k_y))\times\\
&&\displaystyle\hspace*{1cm}\left.\sum_{ \bf k'} (u_{\bf k'}^2 +
v_{\bf k'}^2)2v_{\bf k'}^2 \right]\end{array},\label{eqtp}}
while that from $t^{\prime\prime}$ reads:
\eqn{\begin{array}{ccl}

\delta\epsilon_{\bf k}'' &=& \displaystyle \f{1}{2N} J_3(t^{\prime\prime}) \left[ 2(u_{\bf k}^2 + v_{\bf k}^2) \sum_{\bf k'} 4 v_{\bf k'}^2 \right.\\
&&+ \displaystyle  (u_{\bf k}^2 + v_{\bf k}^2) \sum_{ \bf k'} 2 (u_{\bf k'}^2 + v_{\bf k'}^2) \Big(2\cos(2k_x-2k'_x)\\&&+\hspace*{1cm}\displaystyle 2\cos(2k_y-2k'_y)\Big)\\
&&\displaystyle   +  (u_{\bf k}^2 + v_{\bf k}^2) \sum_{ \bf k'} (u_{\bf k'}^2 + v_{\bf k'}^2)2\Big(\cos(2k'_x)+\cos(2k'_y)\Big)\\
&&\left.\displaystyle +2\Big(\cos(2k_x)+\cos(2k_y)\Big)\sum_{ \bf
k'},(u_{\bf k'}^2 + v_{\bf k'}^2)2v_{\bf k'}^2
\right]\end{array}.\label{eqtpp}}
We now find that the total contribution of the terms coming from
$t$, $t^{\prime}$ and $t^{\prime\prime}$ is no longer independent of wave vector, but
the two-fold degeneracy for each  $\bf k$ remains. The evolution
of $Z_c(\bf k)$ over the Brillouin zone is shown in Fig.
\ref{Zc_fig} for the parameter set (\ref{best-fit}).

Dividing the contributions to $Z_c(\bf k)$
 into three parts coming
from $t$, $t^{\prime}$ and $t^{\prime\prime}$ and denoting an average over the
Brillouin zone by $\langle\cdot\rangle_{\bf k}$ we find
\eqn{\left\{\begin{array}{lcD{.}{.}{4}}
\langle{\Delta_{\bf k}} \rangle& = & 0.187 , \\
\langle{\Delta_{\bf k}'}\rangle & = & 0.019 , \\
\langle{\Delta_{\bf k}''}\rangle & = & 0.013 ,
\end{array}\right.}
using the parameter set (\ref{best-fit}).
Note that here
 $\Delta_{\bf k}'$ and  $\Delta_{\bf k}''$ do
not factorize analytically in Eq.~(\ref{eqtp}) or
(\ref{eqtpp}), rather, for each wave vector, we calculate
$\Delta_{\bf k}'\equiv\delta\epsilon_{\bf k}'/\epsilon_{\bf k}'$ and
similarly for
 $\Delta_{\bf k}''$.
Note also that as $t^{\prime}$ and $t^{\prime\prime}$ become non-zero, even the first
term evolves. This is because $\epsilon_k$ is itself a function of
$t^{\prime}$ and $t"$ and so it changes as these parameters are switched
on.

\begin{figure}
\includegraphics[scale=0.5]{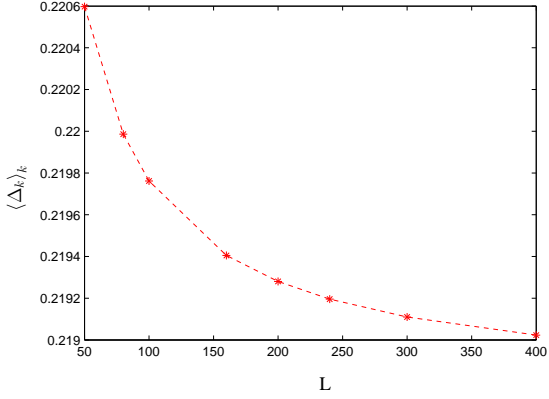}
\caption{(Color online) Evolution of ${\langle \Delta_k \rangle}_k$ as a function
of the number of points in the Brillouin zone}\label{fig_evol}
\end{figure}

\begin{figure}
\includegraphics[scale=0.5]{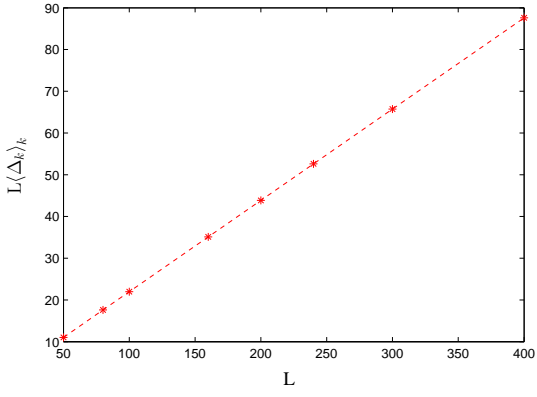}
\caption{(Color online) Evolution of $L\times{\langle \Delta_k \rangle}_k$ as a
function of the number of points in the Brillouin zone.}
\label{fig_evol2}
\end{figure}

As discussed in Section~\ref{Fitting}, incorporating the ${\bf k}$-dependence
of $Z_c({\bf k})$ makes the fitting procedure somewhat
computationally cumbersome and slow from a CPU speed point of view.
There are two reasons for this. Firstly, it is due to the
need to constantly redo the ``internal'' sum over ${\bf k'}$ in
Eqs.~(\ref{eqt}), (\ref{eqtp}) and (\ref{eqtpp}) whenever the $t$, $t^{\prime}$,
$t^{\prime\prime}$ and $U$ parameters are readjusted.
Secondly, to compound this problem the CPU time is further increased by
noting that the sums over ${\bf k'}$ in
Eqs.~(\ref{eqt}), (\ref{eqtp}) and (\ref{eqtpp})
converge somewhat slowly with system size.
${\bf k'}$ in equations (\ref{eqt},\ref{eqtp},\ref{eqtpp})
This is illustrated in Fig.
\ref{fig_evol}, for a magnetic Brillouin zone of linear dimension
$L$, for the parameter set (\ref{best-fit}) (here the lattice
parameter is taken to be unity)
Re-plotting the data in Fig. \ref{fig_evol2}, we show that
${\langle \Delta_k \rangle}$ scales with system size as
 \eqn{{\langle \Delta_k
\rangle}_k(L) = \f{\alpha}{L} + {\langle \Delta_k \rangle}_k
(\infty)~,\label{evo_eq}}
where $\alpha$ is a constant, from which we find:
\eqn{{\langle \Delta_k \rangle}_k (\infty) \,\simeq\, 0.219~.}
The width of the dispersion of $\Delta({\bf k})$ over the
Brillouin zone, $\sigma(\Delta({\bf k}))$, scales in a similar way
and we eventually find
\eqn{\frac{\sigma(\Delta_k)}{{\langle \Delta_k \rangle}_k}
(\infty) \,\simeq\, 0.33 \times 10^{-3}~.}
%
%
%
%
Finite size effects lead to negligible corrections compared to the
least squares fitting procedure, for $L\sim 100$. A magnetic
Brillouin zone of this size was used to make the fits described in
the main text.


\section{Partially Constrained Spin-Wave Fits}
\label{Constrained-Fits}

For a number of compounds, such as Sr$_2$CuO$_2$Cl$_2$,
Bi$_2$Sr$_2$CaCu$_2$O$_{8+\delta}$ (BSCCO, Bi-2212) and
YBa$_2$Cu$_3$O$_{6+\delta}$ (YBCO), values of $t^{\prime}\sim -0.3$ eV,
$t^{\prime\prime}\sim 0.2$ eV, and $U>10$ eV are commonly
used~\cite{Damascelli}. However, these values are not truly
universal among the cuprates. For example, in
Ref.~[\onlinecite{Andersen:2001}], local density approximation (LDA)
calculations do predict variations from one compound to the other.
In particular, for La$_2$CuO$_4$, $t^{\prime}/t =-0.17$ is obtained,
and which corresponds to the value used in ARPES data
analysis~\cite{Ino:2002}.

A recent ARPES experiment by Yoshida {\it et
al.}~\cite{Yoshida:2005} on {\it doped} La$_{2-x}$Sr$_x$CuO$_4$
finds that $t^{\prime}$ varies slightly between compositions. For
$x=0.03$, they obtain $t^{\prime}/t\approx -0.2$ and $t^{\prime\prime}/t^{\prime}\approx 0.5$.
To the best of our knowledge, there exist no ARPES measurements on
undoped La$_2$CuO$_4$, but one may extrapolate to $t^{\prime}/t=-0.21$ for
this composition. In this context, we have performed a constrained
fit to the spin-wave dispersion data of
Ref.~[\onlinecite{coldea01}], imposing $t^{\prime}/t=-0.2$, $t^{\prime\prime}/t^{\prime}=0.5$
and $t=0.25$ eV and only allowing $U$ to vary. We were unable to
find a reasonable value of $U$ with such constraints.
Indeed, we found $U\sim 1.6$ ev with a poor quality of fit.

We then allowed {\it both} $t$ and $U$ to vary. The motivation
being that the effective ratios $t^{\prime}/t=-0.2$ and $t^{\prime\prime}/t^{\prime}=0.5$ could
be well determined by ARPES, while the value of $t$ determined
from the electronic bandwidth could be strongly renormalized. The
best fit then obtained was $t = 0.3430$ eV and $U = 2.55$ eV,
hence $U/t=7.42$. Compared to the fit with free $t$, $t^{\prime}$, $t^{\prime\prime}$
and $U$ values, the overall $\chi^{2}$, with
\begin{eqnarray}
\chi^2 = &\sum_{n}& (E^{\rm experimental}_n - E^{\rm fit}_n)^2 /\\ \nonumber
&&{ ( {\rm experimental\; uncertainty} )}_n       \; ,
\end{eqnarray}
where $E_i$ is $n$'the magnon energy data point, is increased by
roughly 25\% for the fit with $t^{\prime}/t$ and $t^{\prime\prime}/t^{\prime}$ constrained
compared with the fit with
 the values in Eq.~(\ref{best-fit}).

This value of $t\approx 0.34$ eV agrees roughly with those found by
Coldea {\it et al.}~\cite{coldea01}, and the value $U\approx 2.55$
is in-between that of Ref.~[\onlinecite{coldea01} and
the one found in this work and reported in Eq.~(\ref{best-fit}).
In other words, for the
constrained values of $t^{\prime}$ and $t^{\prime\prime}$ used in such a fit, the
values are sufficiently ``small'' that the results for $t^{\prime}=t^{\prime\prime}=0$
of Ref.~[\onlinecite{coldea01}] are more or less recovered.
In other words, the
ratio $t/U$ is a crucial parameter in the fit.
Working with a fit with $t^{\prime}/t$ and $t^{\prime\prime}/t$ fully constrained leads
to a (reduced) ratio $t/U$ compared to the unconstrained  fit of Eq.~(\ref{best-fit}).
Yet, this reduced $U=2.55$ eV
also gives a reduced $t=0.34$ ev,
with the result that while the unconstrained fit has
$t/U=0.126$,
the constrained fit has
$t/U=0.34/2.55=0.133$ (I just calculated) and Coldea {\it et al.}
finds $t/U=0.135$.
Given that $t^{\prime}/t^{\prime\prime}$ is essentially the
same in the constrained and unconstrained fits,
one can think of the constrained fit as
an intermediate point
in switching on $t^{\prime}$ and $t^{\prime\prime}$  at constant $t^{\prime}/t^{\prime\prime}$.
Indeed, we find the following progression of values,
moving from the Coldea {\it et al.} results, the constrained fit results and
the unconstrained fit results:
$U=$ 2.3 eV, 2.55 eV and 3.34 eV,
$t=$ 0.3 eV, 0.34 eV and 0.42 eV, and
$t/U=$ 0.135, 0.133, and 0.126.
This progression of parameters appears monotonous.
Given that we expect
$t^{\prime}$ and $t^{\prime\prime}$ to frustrate N\'eel ordering,
we also expect the metal insulator transition also
to appear for smaller values of $t/U$  with
increasing $t^{\prime}$ and $t^{\prime\prime}$.
Our results are reassuring in
this respect because we find
the fitted $t/U$ value
getting progressively small as $t^{\prime}$ and $t^{\prime\prime}$ are turned on,
consistent with the system remaining in the insulating N\'eel ordered state.

\bibliographystyle{apsrev}
\bibliography{bibli}

\end{document}